\documentclass{aa}
\usepackage[utf8]{inputenc}
\usepackage[usenames,dvipsnames,svgnames,table]{xcolor}
\usepackage{rotating}
\usepackage{subfigure}
\usepackage{graphicx}
\usepackage{txfonts}
\usepackage{color}
\usepackage[none]{hyphenat}

\usepackage[]{hyperref}
\hypersetup{colorlinks=true,citecolor=blue,  linkcolor=blue,urlcolor=black}

\def\Gp{g_{\rm p}}
\def\Mp{M_{\rm p}}
\def\Rp{R_{\rm p}}
\def\Sp{S_{\rm p}}

\def\Mearth{M_\oplus}
\def\Rearth{R_\oplus}
\def\Searth{S_\oplus}

\begin{document}

   \title{Modelling climate diversity, tidal dynamics and the fate of volatiles on TRAPPIST-1 planets}

   \author{Martin Turbet \inst{1}, Emeline Bolmont \inst{2}, Jeremy Leconte \inst{3}, Francois Forget \inst{1}, Franck Selsis \inst{3}, Gabriel Tobie \inst{4}, Anthony Caldas \inst{3}, Joseph Naar \inst{1,5} and Micha\"{e}l Gillon \inst{6}}

 \institute{Laboratoire de M\'et\'eorologie Dynamique, IPSL, Sorbonne Universit\'es, UPMC Univ Paris 06, CNRS, 4 place Jussieu, 75005 Paris, France.
 \email{martin.turbet@lmd.jussieu.fr}
 \and Laboratoire AIM Paris-Saclay, CEA/DRF - CNRS - Universit{\'e} Paris Diderot, IRFU/SAp Centre de Saclay, 91191, Gif-sur-Yvette.
 \and Laboratoire d'astrophysique de Bordeaux, Univ. Bordeaux, CNRS, B18N, all\'ee Geoffroy Saint-Hilaire, 33615 Pessac, France.
 \and Laboratoire de Plan\'etologie et G\'eodynamique, UMR-CNRS 6112, University of Nantes, 2 rue de la Houssini\`ere, F-44322 Nantes, France. 
 \and D\'epartement de g\'eosciences, \'Ecole Normale Sup\'erieure, PSL Research University, 75005 Paris, France.
 \and Space Sciences, Technologies and Astrophysics Research (STAR) Institute, Universit\'e de Li\`ege, All\'ee du 6 Ao\^ut 19C, B-4000 Li\`ege, Belgium.}

   \date{Received; accepted}

\abstract{TRAPPIST-1 planets are invaluable for the study of comparative planetary science outside our Solar System and possibly habitability. 
Both Time Transit Variations (TTV) of the planets and the compact, resonant architecture of the system suggest that TRAPPIST-1 
planets could be endowed with various volatiles today. 
First, we derive from N-body simulations possible planetary evolution scenarios, and show that all the planets are likely in synchronous rotation.
We then use a versatile 3-D Global Climate Model (GCM) to explore the possible climates of cool planets around cool stars, with a focus on the 
TRAPPIST-1 system. We look at the conditions required for cool planets to prevent possible volatile species to be lost 
permanently by surface condensation, irreversible burying or photochemical destruction. We also explore the resilience of the same volatiles 
(when in condensed phase) to a runaway greenhouse process.
We find that background atmospheres made of N$_2$, CO or O$_2$ are rather resistant to atmospheric collapse. However, even if 
TRAPPIST-1 planets were able to sustain a thick background atmosphere by surviving early X/UV radiation and stellar wind atmospheric erosion, 
it is difficult for them to accumulate significant greenhouse gases like CO$_2$, CH$_4$ or NH$_3$. CO$_2$ can easily 
condense on the permanent nightside, forming CO$_2$ ice glaciers that would flow toward the substellar region. 
A complete CO$_2$ ice surface cover is theoretically possible on TRAPPIST-1g and h only, but CO$_2$ ices should be gravitationally 
unstable and get buried beneath the water ice shell in geologically short timescales. Given TRAPPIST-1 planets large EUV irradiation 
(at least $\sim$~10$^3~\times$ Titan's flux), CH$_4$ and NH$_3$ are photodissociated rapidly and are thus hard to accumulate in 
the atmosphere. Photochemical hazes could then sedimentate and form a surface layer of tholins that would progressively thicken over 
the age of the TRAPPIST-1 system.
Regarding habitability, we confirm that few bars of CO$_2$ would suffice to warm the surface of TRAPPIST-1f and g above 
the melting point of water. We also show that TRAPPIST-1e is a remarkable candidate for surface habitability. 
If the planet is today synchronous and abundant in water, then it should always sustain surface liquid water at 
least in the substellar region, whatever the atmosphere considered.
\medskip
\medskip
\medskip
\medskip
\medskip
\medskip
\medskip
\medskip
}

\titlerunning{Modelling climate diversity, tidal dynamics and the fate of volatiles on TRAPPIST-1 planets}
\authorrunning{M. Turbet et al.}

\maketitle

\section{Introduction}


TRAPPIST-1 planets recently discovered by \cite{Gillon2016,Gillon2017} are the closest known transiting temperate Earth-sized exoplanets. The TRAPPIST-1 system hosts at least seven planets that are similar in size (from $\sim$~0.72 to $\sim$~1.13$R_\oplus$) and irradiation (from $\sim$~0.14 to $\sim$~4.3$S_\oplus$) to Solar System rocky planets. Considering that the parent star TRAPPIST-1 is an ultra-cool ($T_{\rm eff}$~=~2550~K), low-mass ($M_{\star}$~=~0.09~M$_{\odot}$) star, the planets of the system should have potentially followed evolutionary pathways very different from what the Solar System planets experienced. They are therefore invaluable probes for comparative planetary science and habitability.

In a first approach, we can speculate on what TRAPPIST-1 planets might look like by comparing their size and irradiation with Solar System planets.
TRAPPIST-1b (0.64~S$_{\text{Mercury}}$) and TRAPPIST-1c (1.19~S$_{\text{Venus}}$) might be airless planets like Mercury, or endowed with a thick atmosphere like Venus.  
TRAPPIST-1d (1.14~$\Searth$) is located near the inner edge of the Habitable Zone, traditionally defined as the range of orbital distances within which a planet can possibly maintain liquid water on its surface \citep{Kasting1993,Kopparapu2013}. Its capacity to host surface oceans, assuming an Earth-like atmosphere and water content, should depend on 1) its rotation mode and 2) subtle cloud albedo feedbacks \citep{Yang2013,Kopparapu2016}. TRAPPIST-1e (0.66~$\Searth$) lies at the right distance to maintain surface liquid water if enough water is available and the atmosphere suitable. TRAPPIST-1f (0.89~S$_{\text{Mars}}$) and TRAPPIST-1g (0.60~S$_{\text{Mars}}$), being slightly bigger than Mars, could have retained a thick CO$_2$ atmosphere and thus conditions potentially similar to Early Mars for a long period of time. Eventually, TRAPPIST-1h (0.30~S$_{\text{Mars}}$, 12 S$_{\text{Titan~/~Enceladus}}$) could look like a warm CH$_4$/N$_2$-rich Titan-like planet, or a Snowball icy-moon-like planet.

However, TRAPPIST-1 planets formed and evolved in a very different environment from our Solar System planets.
At any rate, each of the seven planets may potentially all be airless today, as a result of the star extreme X/UV irradiation \citep{Wheatley:2017,Bourrier2017} and stellar wind through history that could have blown away their atmosphere \citep{Airapetian2017, Dong2017, Garcia-Sage2017, Dong2017b}. Moreover, during the first hundreds of million years following their formation, while TRAPPIST-1 was a pre-main-sequence star and its luminosity significantly higher, each of the seven planets could have faced a runaway phase where the most condensable volatiles (e.g. water) would have been vaporized, and exposed to atmospheric escape. As much as several Earth ocean hydrogen content could have been lost in the process \citep{Bolmont2017,Bourrier2017}. The remaining oxygen could have therefore potentially built up in the atmosphere of the seven TRAPPIST-1 planets \citep{Luger:2015}.

Transit-timing variations (TTVs) measurements of TRAPPIST-1 planets \citep{Gillon2017,Wang:2017} suggest that planet bulk densities are compatible with terrestrial or volatile-rich composition. The latter possibility, and the fact that TRAPPIST-1 planets are in a near-resonant chain, suggest that the planets could have formed far from their star and migrated afterward to their current position. The planets may thus have been formed near or beyond the snowline and could have remained enriched in volatiles until now, despite a potentially massive early atmospheric escape \citep{Bolmont2017}. Uncertainties on the masses derived from TTVs are still affected by significant uncertainties but TTVs will eventually provide robust constraints on the density and volatile content.

Transit spectroscopy with the Hubble Space Telescope (HST) has been done on the two innermost planets, and suggest that they do 
not have a cloud/haze free H$_2$-dominated atmosphere \citep{Dewit2016}. 
This is somehow consistent with the fact that primordial H$_2$ enveloppes 
would have been exposed to efficient atmospheric escape on the small TRAPPIST-1 planets.
At any rate, TRAPPIST-1 planets could still harbor a large variety of atmospheres, such as thick H$_2$O, CO$_2$, N$_2$, O$_2$ or CH$_4$ dominated atmospheres (see the review by \citealt{Forget:2014}). 
In any case, each of these seven planets should be amenable to further characterization by the James Webb Space Telescope (JWST) as early as 2019 \citep{Barstow:2016,Morley2017}. 

\medskip

The goal of the present study is to explore in more details the possible climates of temperate-to-cold planets 
orbiting synchronously around cool stars in general, with a focus on the TRAPPIST-1 system (e,f,g,h). 
The constraints that we derive on their possible atmospheres could serve as a guideline to prepare future observations with JWST.
We explore in this work the conditions required for the coldest TRAPPIST-1 planets to prevent possible volatile species from atmospheric collapse, escape, or photodissociation. 
TRAPPIST-1 is a particular system where even weakly irradiated planets should likely be tidally locked. On synchronously rotating planets, the surface temperature of the cold points can be extremely low, making the different volatile species (N$_2$, CH$_4$, CO$_2$, etc.) highly sensitive to nightside trapping and potentially atmospheric collapse. 

Conversely, we explore the stability of the same volatile species in the condensed phase (either icy or liquid) and on the surface, either on the dayside or the nightside. This condition, widely known for water as the runaway greenhouse limit, is extended here to other molecular species.

Because these processes (runaway and collapse) are 3-D on a synchronous planet, the most suited tools to explore them are 3D Global Climate Models (GCM). 

\medskip

In Section~\ref{lmd_gcm}, we describe our 3D Global Climate Model, and more generally the physical parameterizations adopted in this work. In Section~\ref{tide_section} we discuss the effect of tides on the rotation of TRAPPIST-1 planets. In the next sections, we explore the possible climates that can be expected on the four outer TRAPPIST-1 planets assuming that they are tidally locked and endowed with various volatiles: 
We discuss in Section~\ref{atmosphere_section} the ability of the three outer TRAPPIST-1 planets to sustain an atmosphere of background gases (N$_2$, CO or O$_2$). Then, we explore whether we should expect oxidized CO$_2$-dominated atmosphere (in Section~\ref{co2_section}) or reduced CH$_4$-dominated atmosphere (in Section~\ref{reduced_section}). Eventually, we derive in Section~\ref{trappist_habitability} all the implications for the habitability of TRAPPIST-1 planets.


\section{Method - the LMD Generic Global Climate Model}
\label{lmd_gcm}

The LMD Generic Model is a full 3-Dimensions Global Climate Model (GCM) that initially derives from the LMDz Earth \citep{Hour:06} and Mars \citep{Forget:1999} GCMs . Since then, it has been extensively used to study a broad range of (exo)planetary atmospheres  \citep{Word:11ajl,Forg:13,Word:13,Char:13,Leco:13nat,Leco:13,Word:15,Charnay:2015a,Charnay:2015b,Bolm:16aa,Turb:16,Turb:17icarus,Turb:17epsl}. 

Simulation input parameters include the observed characteristics of TRAPPIST-1 planets \citep{Gillon2017,Luger2017}, 
as summarized in Table~\ref{table_planet}. 
All simulations were performed assuming a circular orbit, 
a choice motivated by the small value of the maximum eccentricities derived from the stability of the system \citep{Gillon2017,Luger2017}. 
Even for a non circular orbit, the orbital period is sufficiently small that the eccentricity should probably be quite high to significantly impact the climate of synchronous planets (see \citealt{Bolm:16aa} for their 10$^{-4}$ L$_\text{sun}$ case). We assumed that each of the planets is in synchronous rotation with 0$^{\circ}$ obliquity, as supported by calculations presented in Section~\ref{tide_section}.

\begin{table*}
\centering
\caption{Adopted planetary characteristics of TRAPPIST-1 planets for climate simulations.}
\begin{tabular}{lcccccccc}
\hline
Parameter & Tb & Tc & Td & Te & Tf & Tg & Th & Unit \\
\hline
$\Rp$   & 1.09$^a$ & 1.06$^a$ & 0.77$^a$ & 0.92$^a$ & 1.05$^a$ & 1.13$^a$ & 0.75$^b$ & $\Rearth$     \\
$\Mp$   & 0.85$^a$ & 1.38$^a$ & 0.41$^a$ & 0.62$^a$ & 0.68$^a$ & 1.34$^a$ & 0.38 (arb.) &  $\Mearth$\\
$\Gp$   & 7.07 & 12.14 & 6.75 & 7.22 & 6.11 & 10.34 & 6.6 (arb.) & m~s$^{-2}$\\
Semi-major axis   & 0.011$^a$ & 0.015$^a$ & 0.021$^a$ & 0.028$^a$ & 0.037$^a$ & 0.045$^a$ & 0.060$^b$ & au      \\
$\Sp$     & 4.25$^a$ & 2.27$^a$ & 1.143$^a$ & 0.662$^a$ & 0.382$^a$ & 0.258$^a$ & 0.165$^b$ & $\Searth$     \\
$\Sp$     & 5806 & 3101 & 1561 & 904 & 522 & 352 & 225 & W~m$^{-2}$     \\
Spin-orbit resonance         & &   &      & 1:1 & & &   \\
Period          & 1.51$^a$ & 2.42$^a$ & 4.05$^a$ & 6.10$^a$ & 9.21$^a$ & 12.35$^a$ & 18.76$^b$ & Earth days \\
$\Omega_p$  & 4.82 & 3.00 & 1.80 & 1.19 & 0.790 & 0.589 & 0.388 & 10$^{-5}$ rad~s$^{-1}$ \\
Obliquity         & & & & 0 & & & & $^{\circ}$\\
Eccentricity         & & & & 0 & & &   \\
\hline
\end{tabular} 
\tablefoot{Most of the values derive from \citet{Gillon2017}$^a$ and \citet{Luger2017}$^b$. Note that mass estimates are all compatible (at less than 1 $\sigma$) with the TTV analysis of \citet{Wang:2017} that included both Spitzer \citep{Gillon2017} and K2 \citep{Luger2017} transits .}
\label{table_planet} 
\end{table*}

The numerical simulations presented in this paper were all carried out at a horizontal resolution of 64~$\times$~48 (e.g., 5.6~$^{\circ}~\times$~3.8$^{\circ}$) in longitude~$\times$~latitude. 
In all the simulations, the dynamical time step is set to 90~s. The physical parameterizations and the radiative 
transfer are calculated every 15~min and 1~h, respectively.
Subgrid-scale dynamical processes (turbulent mixing and convection) were parameterized as in \citet{Forg:13} and
\citet{Word:13}. The planetary boundary layer was accounted for by the \citet{Mellor:1982} and \citet{Galperin:1988} 
time-dependent 2.5-level closure scheme, and complemented by a convective adjustment which rapidly mixes the
atmosphere in the case of unstable temperature profiles.
A filter is applied at high latitude to deal with the singularity in the grid at the pole \citep{Forget:1999}.
In the vertical direction, the model is composed of 26 distinct atmospheric layers that were built using hybrid $\sigma$ coordinates and 18 soil layers. 
These 18 layers are designed to represent either a rocky ground (thermal inertia I$_{\text{rock}}$~=~1000~J~m$^{-2}$~K$^{-1}$~s$^{-\frac{1}{2}}$), 
an icy ground (I$_{\text{ice}}$~=~2000~J~m$^{-2}$~K$^{-1}$~s$^{-\frac{1}{2}}$) 
or an ocean (I$_{\text{ocean}}$~=~20000~J~m$^{-2}$~K$^{-1}$~s$^{-\frac{1}{2}}$ to take into account the efficient vertical mixing in the first 
tens of meter of the ocean, as previously done in \citet{Leco:13nat} and \citet{Char:13}) 
depending on the assumed surface. Since all the simulations were carried out for a synchronous rotation, thermal inertia should only affect the variability of the atmosphere. Oceanic heat transport is not included in this study. 

The GCM includes an up-to-date generalized radiative transfer that takes into account the absorption and scattering 
by the atmosphere, the clouds and the surface from visible to far-infrared wavelengths, as described in \citet{Word:11ajl}.
The radiative transfer is performed here for variable gaseous atmospheric compositions made of various cocktails 
of CO$_{2}$, CH$_4$, N$_2$ and H$_{2}$O, using the correlated-k method \citep{Fu1992,Eymet2016}. Molecular absorption lines were taken 
from HITRAN 2012 \citep{Rothman2012}. Sublorentzian profiles \citep{Perrin:1989,Campargue:2012}, Collision Induced Absorptions 
\citep{Gruszka:1997,Baranov:2004,Wordsworth2010,Richard2012} and various other continua \citep{Gruszka:1997,Clough:05,Richard2012} 
were properly included in the calculations when needed. 
For the computation, we used between 32 and 38 spectral bands in the thermal
infrared and between 36 and 41 spectral bands in the visible domain, depending on the atmospheric composition considered. 
16 non-regularly spaced grid points were used for the g-space integration, where g is the cumulative 
distribution function of the absorption data for each band.
We used a two-stream scheme \citep{Toon:1989} to take into account the scattering effects of the atmosphere and the clouds, 
using the method of \citet{Hansen:1974}.

The emission spectrum of TRAPPIST-1 was computed using the synthetic BT-Settl spectrum\footnote{Downloaded from  https://phoenix.ens-lyon.fr} 
\citep{Rajpurohit2015} assuming a temperature of 2500~K, a surface gravity of 10$^3$~m~s$^{-2}$ and a metallicity of 0~dex.

The GCM directly computes the wavelength-dependent albedo of water 
ice~/~snow from a simplified albedo spectral law of ice~/~snow, calibrated to 
get ice~/~snow bolometric albedo of 0.55 around a Sun-like star, 
as in \citet{Turb:16}. Around TRAPPIST-1, we calculate that the average bolometric albedo for water ice~/~snow is $\sim$~0.21. 
Around an ultra-cool star like TRAPPIST-1, the bolometric albedo of water ice~/~snow is drastically 
reduced \citep{Joshi2012,VonParis_b:2013,Shields:2013} due to the shape of its 
reflectance spectrum \citep{Warren1980,Warren1984}.

Melting, freezing, condensation, evaporation, sublimation, and precipitation of H$_2$O are included in the model. 
Similarly, we take into account the possible condensation/sublimation of CO$_2$ in the atmosphere (and on the surface) when needed but not the radiative effect of CO$_2$ ice clouds because their scattering greenhouse effect \citep{Forg:97} should be low around cool stars like TRAPPIST-1 \citep{Kitzmann:2017} and limited by partial cloud coverage \citep{Forg:13}.
The effect of latent heat is properly taken into account when H$_2$O and/or CO$_2$ condense, evaporate or sublimate.

CO$_2$ and H$_2$O cloud particle sizes are determined from the amount of condensed material and
the number density of cloud condensation nuclei [CCN]. The latter parameter was taken to be constant everywhere in the atmosphere, and 
equal to 10$^6$~kg$^{-1}$ for liquid water clouds, 
10$^4$~kg$^{-1}$ for water ice clouds \citep{Leco:13nat} and 10$^5$~kg$^{-1}$ for CO$_2$ ice clouds \citep{Forg:13}.
Ice particles and liquid droplets are sedimented following a Stokes law described in \citet{Rossow:1978}.
H$_2$O precipitation are computed with the scheme from \citet{Boucher:1995}, with precipitation
evaporation also taken into account.

All the numerical climate simulations were run long enough (up to 30 Earth years) to reach equilibrium. 
Simulations that lead to unstable CO$_2$ surface collapse were 
stopped when the rate of CO$_2$ surface condensation reached a positive constant, as in \citet{Turb:17epsl}. 

Note that more details on the LMD Generic model can be found in \citet{Forget:1999}, \citet{Word:11ajl}, 
\citet{Forg:13}, \citet{Word:13}, \citet{Char:13}, \citet{Leco:13nat}, \citet{Turb:16} and \citet{Turb:17icarus}.




\section{Effect of tides on TRAPPIST-1 planets}
\label{tide_section}

All observed TRAPPIST-1 planets are inside an orbital distance of 0.06~au. As a comparison, Mercury orbits at $\sim$~0.4~au from the Sun. 
For such close-in planets, tidal interactions are expected to be strong and influence the orbital and rotational dynamics of the system.

We use here a standard equilibrium tide model \citep{Mignard1979, Hut1981, EKH1998, Bolmont2011} to estimate the tidal evolution of the system.
We combine an approach based on evolution timescale calculations and N-body simulations of the system using Mercury-T \citep{Bolmont2015}.

Mercury-T is a N-body code, which computes the orbital and rotational evolution of multi-planet systems taking into account tidal forces and their resulting torques (in the equilibrium tide framework), the correction for general relativity and the rotational flattening forces and torques.
From this code, the evolution of the orbital parameters can be calculated (such as semi-major axis, eccentricity and inclination) as well as the rotation of the different bodies (i.e., the rotation period and obliquity of the planets).
This code has previously been used to study the orbital dynamics of close-in and/or compact and/or near-resonant systems 
such as 55-Cnc \citep{Bolmont2013}, Kepler-62 \citep{Bolmont2015} and Kepler-186 \citep{Bolmont2014}, and is now used here for 
the TRAPPIST-1 system.

\subsection{Tidal dissipation and orders of magnitude}

Simple order of magnitude calculations allow us to determine that the tide raised by the planets in the star is a priori negligible for this system today.
Even considering a relatively high dissipation for a purely convective body (i.e., the dissipation of a hot Jupiter as estimated by \citealt{Hansen2010}), we find semi-major axis and eccentricity evolution timescales of $10^8$~Myr and $5\times 10^7$~Myr respectively.
Note that the dissipation is a measure
of how fast the system is tidally evolving: the higher the dissipation, the faster the evolution.
The age of TRAPPIST-1 has recently been estimated to be between 5 and 10~Gyr \citep{Luger2017,Burgasser2017}.
The evolution timescales for semi-major axis and eccentricity are thus consistent with \citet{Bolmont2011}, which showed that the stellar-tide driven evolution around low mass stars and brown dwarfs was negligible for ages superior to $\sim 100$~Myr due to the decrease of the stellar radius.

The system therefore principally evolves due to the gravitational tide raised by the star in the planets (the planetary tide).
The planetary tide mainly acts to decrease the obliquity of the planet, 
synchronize the rotation and on longer timescales decrease the eccentricity 
and semi-major axis. The dissipation in the planets depends on their internal structure and 
thermal state \citep{Henning2014}, as well as on the extension of the external fluid envelop 
(presence of surficial liquids - water ocean, magma ocean - and of a massive atmosphere) 
\citep[e.g.][]{Dermott1979,Remus2015}. On Earth, the dissipation is dominated by the dynamical 
response of the ocean and friction processes along the costline and to a lesser extent to interactions 
with seafloor topography in deep ocean \citep{Egbert2001,Egbert2004,Williams2014}. Dissipation on the Earth 
is highly dependent on the continent configuration and is therefore expected to significantly change on geological 
timescale as a consequence of tectonic plate motion \citep[e.g.][]{Poliakow2005,Williams2014}. The Earth's dissipation 
is close to its highest value right now, and could have varied by a factor of almost ten during the last 200 Myr \citep{Poliakow2005}.\\

In order to take into account the huge uncertainties in the dissipation factors of exoplanets (for which we do not know the internal structure), we consider various dissipation factors for the planets (from 0.01 to 10 times the dissipation of the Earth).
The lowest value we consider here, which is comparable to the dissipation estimated in Saturn \citep{Lainey2017}, would be representative of planets dominated by tidal response of a massive fluid envelop. 
The highest value is close to the maximal possible value and would be representative of very hot planets dominated by fluid-solid friction. 
There is no example in the Solar system of such a dissipative object. 
Even  the highly dissipative Jupiter's moon Io \citep{Lainey2009} has a dissipation function smaller than this extreme value, which is comparable to the Earth's value (even if the dissipation process is very different). 
However, we could envision that Earth-sized bodies with a dissipation process comparable to that of Io could reach such a highly dissipative state. 
The tidal dissipation is also sensitive to the forcing frequency \citep[e.g.][]{Sotin2009,Henning2014}, and therefore to the distance from the star. 
For simplicity, we ignore this effect here and consider constant dissipation functions, independently of the distance from the star and the size of the planet, which is sufficient at first order to provide some typical tendencies.

Considering the dissipation for the planets of the system to be a tenth of the dissipation of the Earth \citep{DeSurgyLaskar1997, Williams2014}, comparable to the dissipation in Mars for instance \citep{Yoder2003,Bills2005}, we find evolution timescales for the rotation to range from $10^{-4}$~Myr for TRAPPIST-1b to 7~Myr for TRAPPIST-1h. 
For the obliquity, the evolution timescales range from $10^{-3}$~Myr for planet-b to 80~Myr for planet-h.
Given the estimated age of the system, all planets are thus expected to have a small obliquity and to be near synchronization.

In the tidal framework we use here, the rotation of the planets tend to pseudo-synchronization if the orbit is not circular. 
However, \citet{Makarov2013} showed that considering a more physical rheology for the planet rather lead to a succession of spin-orbit resonance captures as the eccentricity of the planet decreases.
We discuss the possibility of capture in spin-orbit resonant configuration in the following section.

\medskip

\medskip

\subsection{Should we expect TRAPPIST-1 planets to be all tidally locked?}

\begin{figure*}
    \centering
\includegraphics[width=\linewidth]{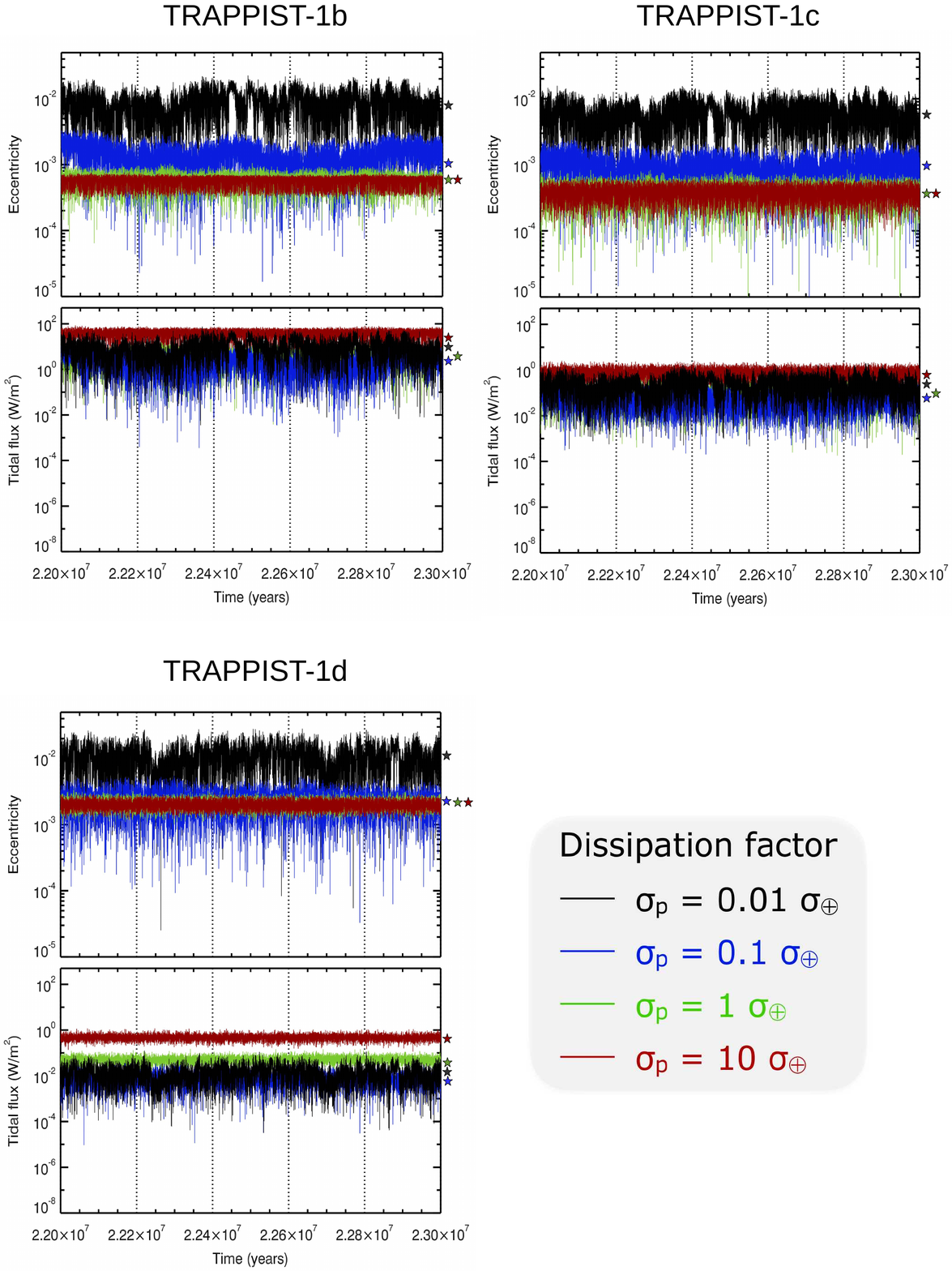}
\caption{Eccentricity (top panels) and tidal heat flux (bottom panels) for the three inner planets of TRAPPIST-1 for different tidal dissipation factors (different colors): from 0.01 to 10 times the Earth's value \citep[taken from][]{DeSurgyLaskar1997}. Stars indicate the mean values.}
\label{ecc_flux_inner_planets}
\end{figure*}

\begin{figure*}
    \centering
\includegraphics[width=\linewidth]{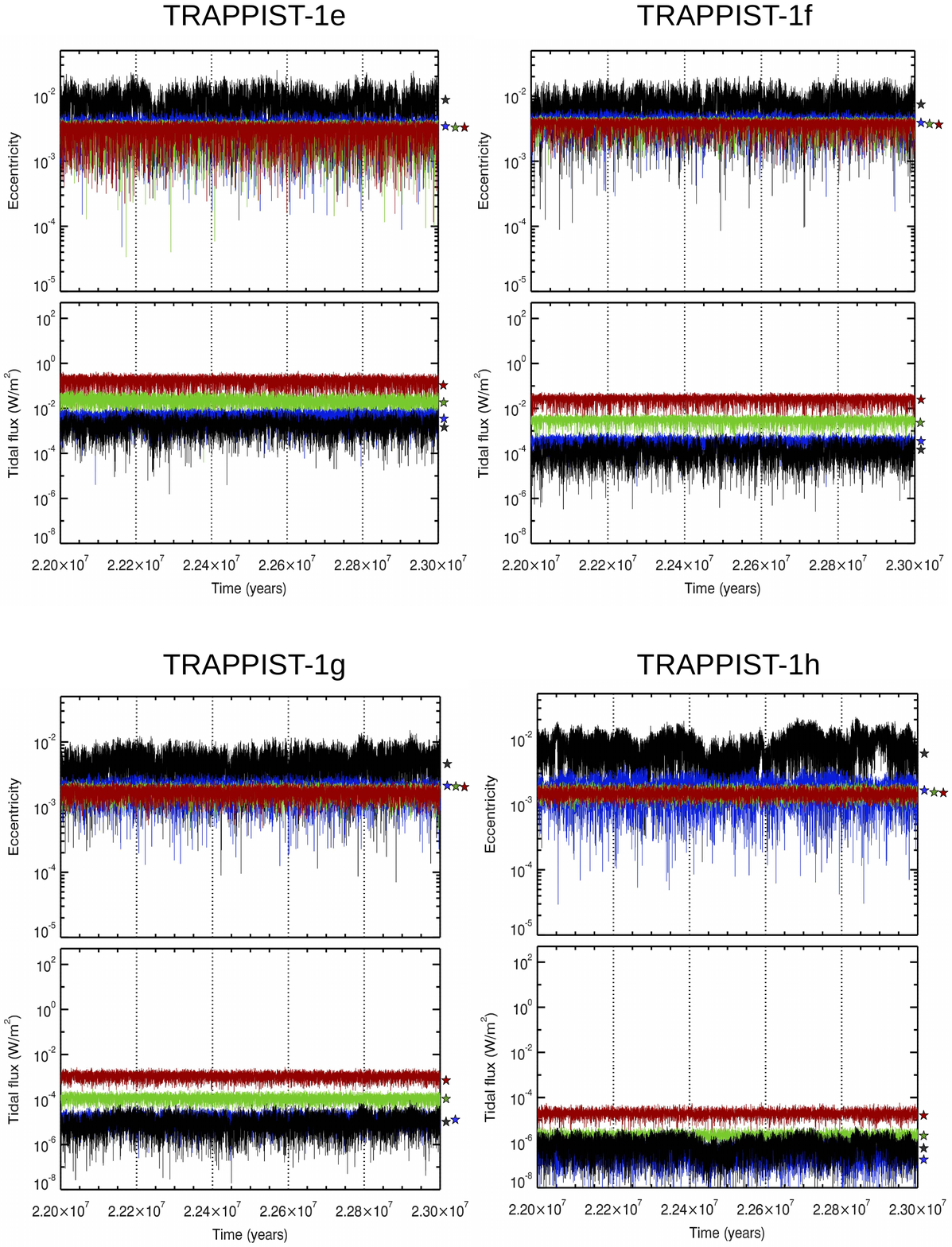}
\caption{Same than Figure~\ref{ecc_flux_inner_planets} but for the four outer planets.}
\label{ecc_flux_outer_planets}
\end{figure*}

\begin{table*}
\centering
\caption{Possible tidal heat flux for TRAPPIST-1 planets coming from a dynamical simulation of the system.}
\begin{tabular}{lcccccccc}
\hline
Parameter & Tb & Tc & Td & Te & Tf & Tg & Th & Unit \\
\hline
ecc mean ($\times 10^{-3}$) & $0.6$ & $0.5$ & $3.9$ & $7.0$ & $8.4$ & $3.8$ & 2.8 & \\
ecc max ($\times 10^{-3}$) & $1.5$ & $1.2$ & $5.9$ & $8.3$ & $9.7$ & $4.8$ & 4.0 & \\
$\Phi_{\rm tid}$mean& 4.8   & 0.17 & 0.17 & 0.09 & 0.01 & $<10^{-3}$ & $<10^{-4}$  & W~m$^{-2}$\\
$\Phi_{\rm tid}$max & 25    & 0.90 & 0.38 & 0.12 & 0.02 & $<10^{-3}$ & $<10^{-4}$  & W~m$^{-2}$\\
\hline
\end{tabular} 
\label{table_flux} 
\end{table*}

With such short period orbits, it is often assumed that bodily tides have spun-down the planets to the spin-orbit synchronous resonance in a relatively short time. However, it is now known that some other processes can sometimes act to avoid the synchronous state. We will thus briefly review these processes. However, it appears that around such a low mass star, none of them is strong enough to counteract bodily tides so that all TRAPPIST-1 planets are probably in a synchronous-rotation state.

Indeed, one of the possibility for planets on an eccentric orbit is the capture into a higher order spin-orbit resonance \citep{GP66}. 
However, as discussed by \citet{Ribas2016} for the case of Proxima Centauri~b, around a low mass star, the question is whether the dissipative tidal torque exerted by the star on the planet is strong enough to avoid the capture into resonance (which is permited by the non-axisymetric deformation of the planet). 
We use the methods detailed in the section~4.6 of \citet{Ribas2016} to calculate the probability of spin-orbit resonance capture of the planets of TRAPPIST-1.
This method relies on comparing the tidal torque and the triaxiality torque, which depend strongly on eccentricity.
The lower the eccentricity, the lower the spin-orbit resonance capture probability. 
For the capture to be possible, the eccentricity of a given planet in the system would need to be roughly above 0.01. The capture probability becomes greater than 10\% only for an eccentricity greater than 0.03. However, simulations of the dynamics of the system accounting for tides and planet-planet interactions (see below) seem to show that such eccentricities are on the very high end of the possible scenarios. The spin orbit capture is thus seen as rather improbable in such a compact system.

The other possibility is that thermal tides in the atmosphere can create a strong enough torque to balance the stellar tidal torque on the mantle, as is expected to be the case on Venus \citep{leconte2015,Auclair-Desrotour2016}. For this process to be efficient, the planet must be close enough from the star so that tides in general are able to affect the planetary spin, but far enough so that bodily tides are not strong enough to overpower atmospheric tides. In a system around such a low-mass star, this zone rests well beyond the position of the seven discovered planets (see Fig~3 of \citealt{leconte2015}). Atmospheric tides are thus unable to affect the spin of the planet significantly.

\subsection{Tidal N-body simulations}

We then performed N-body simulations using Mercury-T \citep{Bolmont2015} to compute the complete evolution of the system, taking into account tides, general relativity and the rotational flattening of the different rotating bodies.
We explored the dissipation factors range discussed above (from 0.01 to 10 times the dissipation of the Earth).
Figures~\ref{ecc_flux_inner_planets} and \ref{ecc_flux_outer_planets} show the evolution of the eccentricity and resulting tidal heat flux for the different planets of TRAPPIST-1 and for the different dissipation factors.

The initial state of our simulations corresponds to the orbital state of the system determined in \citet{Gillon2017} for planets b to g, and we used \citet{Luger2017} for the orbital parameters of planet h.
As the evolution timescales of rotation and obliquity are small compared to the estimated age of the system, we considered the planets to be initially in synchronization and with a very small obliquity.
We considered two sets of initial eccentricities: all eccentricities at $10^{-6}$ and the eccentricities derived from TTVs of \citet{Gillon2017}.

All simulations display the same behavior: after a short initial phase of eccentricity excitation, all excentricities decrease on a timescale depending on the dissipation factor, to reach a mean equilibrium value.
This equilibrium value is the result of the competition between tidal damping and planet-planet excitations \citep[e.g.][]{Bolmont2013} and the eccentricity oscillates around it.
The eccentricities corresponding to the equilibrium value are all relatively small as they are inferior to $10^{-3}$ for planets b and c, and inferior to $10^{-2}$ for planets d to h. 
The equilibrium value depends slightly on the dissipation factor of the planets: the higher the dissipation factor, the smaller the eccentricity. 
Note that such small eccentricities would have no effect on the climate (e.g., \citealt{Bolm:16aa}), that is why we assumed circular orbits for all planets in the climate simulations performed in this study.

The same kind of behavior can be seen for the obliquity of the planets: they assume an equilibrium value, result of the competition between tidal damping and planet-planet excitations. 
The equilibrium values are very small. 
For instance, the obliquities of the planets are smaller than 1$^\circ$, which is why here we also assumed a zero obliquity for all planets.

\subsubsection*{Estimates of internal heat fluxes}

Due to planet-planet interactions, the eccentricities and obliquities of the planets are not zero.
This means that the planets are constantly submitted to a changing potential and constantly being deformed. This implies that the planets get tidally heated.

\begin{figure}
\centering
\centerline{\includegraphics[scale=0.8]{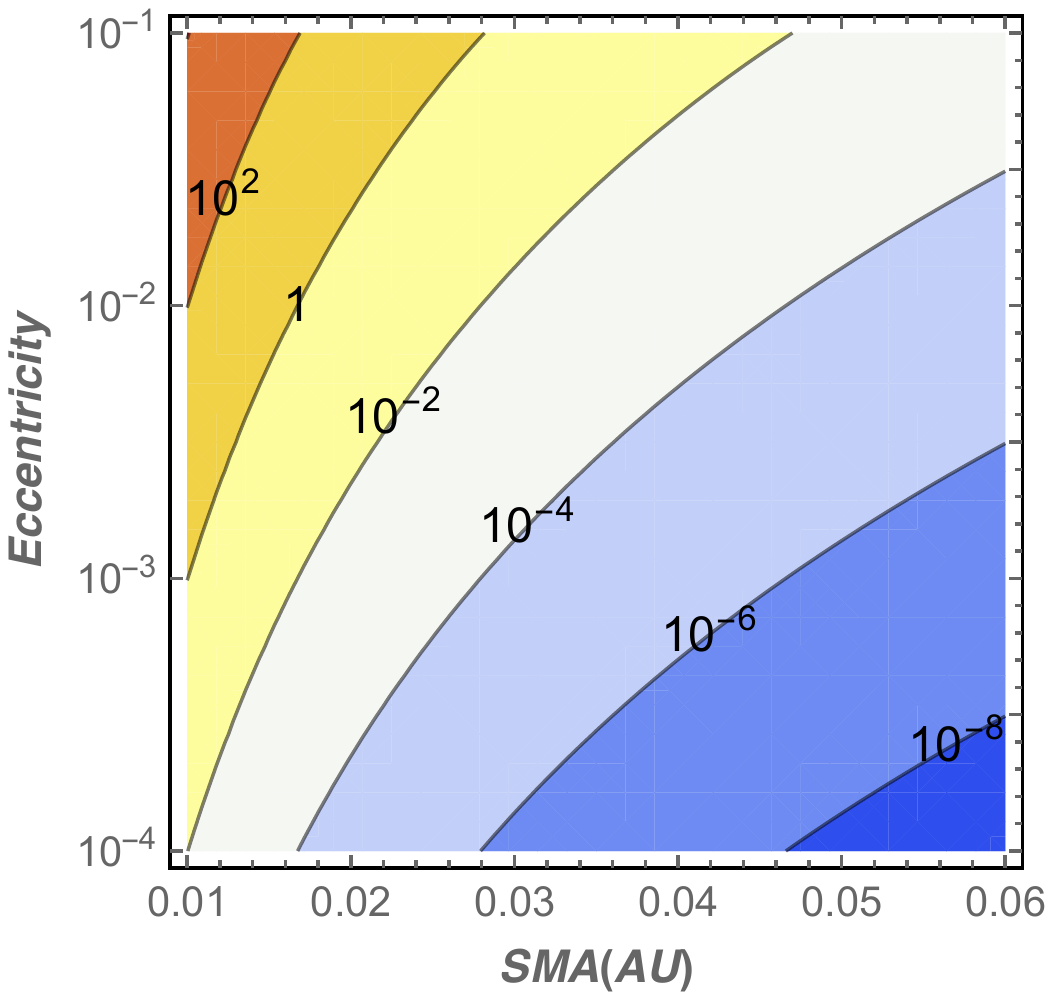}}
\caption{Tidal heat flux map (in W~m$^{-2}$) as a function of semi-major axis (X axis) and eccentricity (Y axis) for planets with the Earth mass and radius. The dissipation efficiency is assumed to be one tenth of the Earth one to account for the dissipation in the mantle only. The tidal flux scales linearly with this parameter. But one has to keep in mind that this value can easily change by orders of magnitude with the internal structure of each planet. This map should thus serve as a rough guide only. }
\label{tidal_heat_flux_map}
\end{figure}

Our simulations with Mercury-T allow us to derive a possible state of the system and the corresponding tidal heat flux for all planets (see Fig~\ref{tidal_heat_flux_map}).
We find that the equilibrium eccentricity is enough to create a significant heat flux for the inner planets.
For instance, assuming the tidal dissipation of the Earth for all the TRAPPIST-1 planets, 
we find that the eccentricity of planet b varies from $\sim 8\times 10^{-6}$ to $1.5 \times 10^{-3}$, with a mean value at $6 \times 10^{-4}$.
These eccentricities lead to a tidal heat flux which varies from $\sim 0.02$~W~m$^{-2}$ to $\sim 25$~W~m$^{-2}$, with a mean value at $\sim 5$~W~m$^{-2}$.
Table~\ref{table_flux} shows possible values for the tidal heat flux for all the planets. 
We warn the reader that the mechanism of electromagnetic induction heating recently proposed by \citet{Kislyakova:2017} should have a negligible 
contribution to the surface heat flux. We calculate from \citet{Kislyakova:2017} (Table 1) that the induction heating should not produce more 
than 8, 19, 8, 0.6, 0.2, 0.08 and 0.03~mW~m$^{-2}$ for TRAPPIST-1b, c, d, e, f, g and h, respectively.
Figures~\ref{ecc_flux_inner_planets} and \ref{ecc_flux_outer_planets} show a snapshot  of the evolution of the eccentricity and tidal heat flux over 1~Myr for each planet and for 4 different tidal dissipation factors.


For the four outer planets of the TRAPPIST-1 system, the mean tidal heat fluxes derived from N-body calculations are lower than 0.09~W~m$^{-2}$, which corresponds roughly to the Earth mean geothermal heat flux \citep{Davies:2010}. Therefore, tidal heat flux is expected to play a minor role on the climate of these planets. It could nonetheless contribute significantly to: 
\begin{enumerate}
    \item the surface temperature of the cold traps, and hence to the atmospheric condensation of background gas (like N$_2$).
    \item the maximum amount of the various sort of volatiles that could be trapped on the cold points of the planets.
    \item the maximum depth of a subsurface liquid water ocean \citep{Luger2017}.
    \item more generally, the internal structure and the orbital dynamics of the planets.
\end{enumerate}
The two first effects are explored in the next sections.

\section{Could TRAPPIST-1 planets be airless planets?}
\label{atmosphere_section}

\begin{figure*}
\centering
\centerline{\includegraphics[scale=0.4]{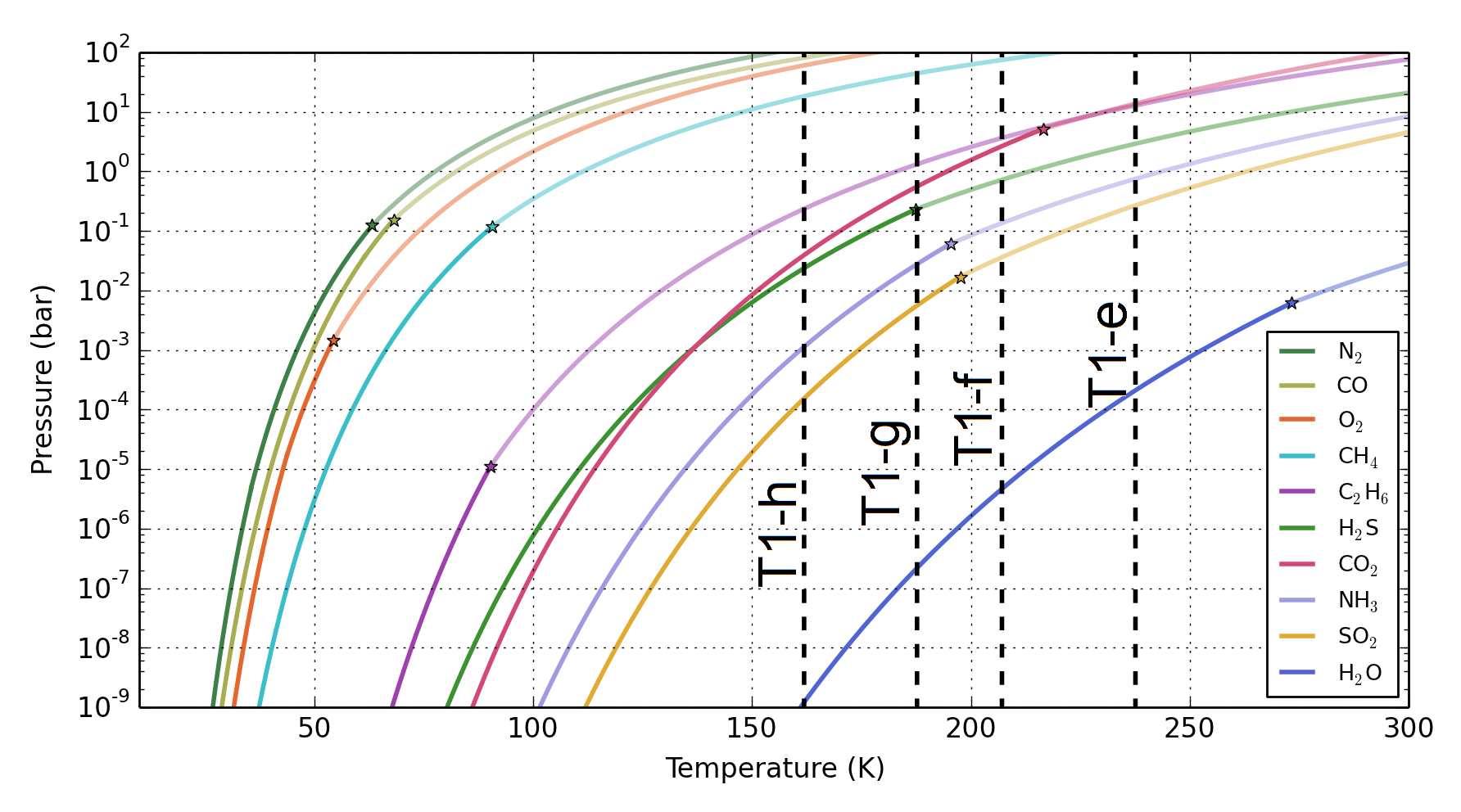}}
\caption{Equilibrium vapor pressures as a function of temperature for 9 different species (where experimental data are available) that could be abundant on TRAPPIST-1 planets. Solid black lines where superimposed to indicate the equilibrium temperatures of TRAPPIST-1e,f,g and h (assuming a surface albedo of 0.2). Stars indicate the positions of the triple points. These curves were adapted from \cite{Fray2009}.}
\label{diagram_thermo}
\end{figure*}


Leaving aside the case of H$_2$/He-rich atmospheres, we focus here specifically on the case of the next three most 
volatile species: N$_2$, CO and O$_2$, because they are the best compromise between volatility 
(see Figure~\ref{diagram_thermo}) and abundance. The cases of CO$_2$ and CH$_4$, which are significantly 
less volatile, are discussed later on in Sections~\ref{co2_section} and \ref{reduced_section}, respectively. 
The main goal of this section is to assess the necessary conditions for the three outer TRAPPIST-1 
planets to sustain a global, background atmosphere (i.e. a rather transparent atmosphere 
that can ensure the transport of heat and the pressure 
broadening of absorption lines of greenhouse gases). 
Background gases are essential because they can prevent the more volatile species such as CO$_2$ or NH$_3$ from collapsing on the nightside.

We assume for now that the surface is covered by water (liquid or icy) because it is expected to be the most abundant volatile, as well as the less dense (this is a key element for planetary differentiation) and most condensable (see Figure~\ref{diagram_thermo}).

We refer the reader to the review by \citet{Forget:2014} (and the references therein) for more information on possible sources and sinks of these volatile species.

\subsection{Can a global atmosphere avoid atmospheric collapse?}
\label{global_atmosphere}

To begin, we assume that the three outer TRAPPIST-1 planets initially start with an atmosphere. On synchronously rotating planets, the nightside surface temperature can be so low that the atmosphere itself starts to condense on the surface. We look for the minimal atmospheric pressure necessary to prevent them from atmospheric collapse, a configuration for which all the volatiles are permanently frozen on the nightside. 

\begin{figure*}
\centering
\centerline{\includegraphics[scale=0.5]{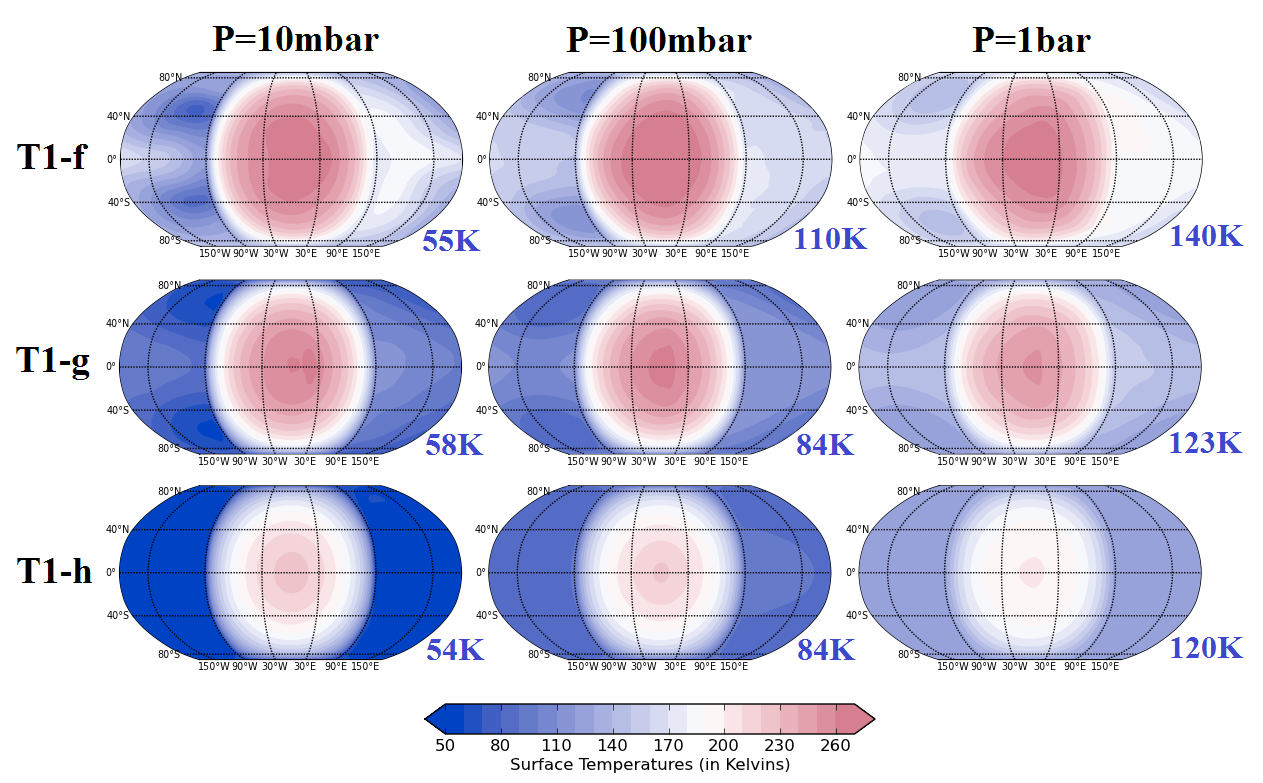}}
\caption{Maps of surface temperatures (averaged over 50 Earth days) for TRAPPIST-1f, g and h, assuming initially cold,
water ice covered frozen planets, 
endowed with a pure N$_2$ atmosphere (with H$_2$O as a variable gas) 
at 3 different surface pressure (10~millibar, 100~millibar, 1~bar). 
Blue colored label indicates the mininum temperature reached by the coldest point of the planet 
throughout the entire simulation.
As a reminder, for partial pressures of 10~millibar (resp. 100~millibar and 1bar), 
N$_2$ is expected to collapse at 53K (resp. 62K and 79K), CO at 56K (resp. 66K and 83K), 
and O$_2$ at 61K (resp. 72K and 87K). The geothermal heat flux is not taken into account, 
but the coldest temperatures found in these cases a posteriori show that it can be neglected.}
\label{water_rich_temperature_map}
\end{figure*}

For this, we performed several simulations of TRAPPIST-1f, g and h planets (surface albedo fixed to 0.2 corresponding to a water ice surface around TRAPPIST-1, or coincidentally to a rocky surface) endowed with a pure N$_2$ atmosphere (with H$_2$O as a variable species) for various atmospheric pressures (from 1~bar down to 10~millibar). Surface temperature maps corresponding to these experiments are shown in Figure~\ref{water_rich_temperature_map}. 

We find that a pure N$_2$ atmosphere (with H$_2$O as a variable gas) is quite resistant to atmospheric collapse for each of the three TRAPPIST-1 (fgh) outer planets. A collapse would be expected for N$_2$ partial pressure (pN$_2$) slightly lower than 10~millibar, and this value should hold for each of the 3 planets notwithstanding their various levels of irradiation. Our simulations indicate in fact (see Fig~\ref{water_rich_temperature_map}) that if TRAPPIST-1h is always globally colder than TRAPPIST-1g (which is globally colder than planet f), it is not necessarily the case for the temperature of their cold points. 
TRAPPIST-1f, g and h planets have rotation periods $\sim$~10$^1$~Earth days and they lie thus near 
the transition between slow and fast rotating regimes \citep{Edson2011,Carone15,Carone16}. 
They should be in one of these two regimes and could potentially be in both, depending on the initial 
forcing \citep{Edson2011}. Since the temperature of the cold points is critically dependant on the circulation 
regime (see \citealt{Carone16}, their Figure~1,2,3), it is difficult to assess which of these 3 TRAPPIST-1 
planets should be more sensitive to atmospheric collapse.


In the same fashion than N$_2$, CO and O$_2$ are rather transparent in the infrared region of the surface thermal emission 
(between 10 and 100~microns, here) and have a similar molar mass than N$_2$ (between 28 and 32~g~mol$^{-1}$). 
We can then safely extend our results for N$_2$ to CO and O$_2$-dominated atmospheres. 
These two gases are slightly more condensable gas and are thus expected to collapse for atmospheric pressure 
slightly higher than 10~millibar (see the legend of Figure~\ref{water_rich_temperature_map}). 
These results could be tested in future studies with models that would properly take into account 
the radiative properties of a CO or O$_2$-dominated atmosphere, and that would explore the sensitivity 
of these results to the assumption made on the surface composition (water, rock, etc).

More generally, note that as the dominant gas becomes less and less volatile, building up an atmosphere becomes more and more complicated due to atmospheric collapse. At any rate, such collapse would trigger a positive feedback, because as the atmosphere condenses, the heat redistribution would become less efficient, leading to even more condensation. This would drive the planets to a complete and irreversible atmospheric collapse. 

\subsection{How much volatile can be trapped on the nightside of an airless planet?}

Conversely, we suppose now that the planets initially start without a global atmosphere, which could have been blown away during the early active phase of TRAPPIST-1. In this configuration, all the volatiles (accreted, outgassed, or residual from the initial collapse) are expected to accumulate on the cold side of the planet. We calculate here the maximum amount of volatiles that could be trapped in ice caps before a global atmosphere would be (re)formed.

The nightside surface temperature T$_\text{night}$ on an airless tidally locked planet is determined by the geothermal heat flux F$_{\text{geo}}$ of the planet:
\begin{equation}
T_\text{night} = \left(\frac{F_{\text{geo}}}{\sigma}\right) ^{\frac{1}{4}}.
\label{eq_ground_collapse}
\end{equation}
For geothermal heat flux of 500~mW~m$^{-2}$  (corresponding to planets with strong  tidal dissipation in their interior) 
(resp. 50~mW~m$^{-2}$), we get a nightside surface temperature of 50~K (resp. 30~K). 
The temperature at the base of the nightside ice cap (e.g. the temperature below the volatiles) 
depends on the geothermal heat flux $F_{\text{geo}}$ and the 
nightside surface temperature $T_\text{night}$. When the ice cap is full 
(e.g. when ices start to convert into liquids), the temperature at the base of the 
nightside ice cap should be close (always higher, though) to that of the triple point. 
The triple point temperature is equal to 63K for N$_2$, 68K for CO, and 55K for O$_2$.

\begin{figure}
\centering
\centerline{\includegraphics[scale=0.5]{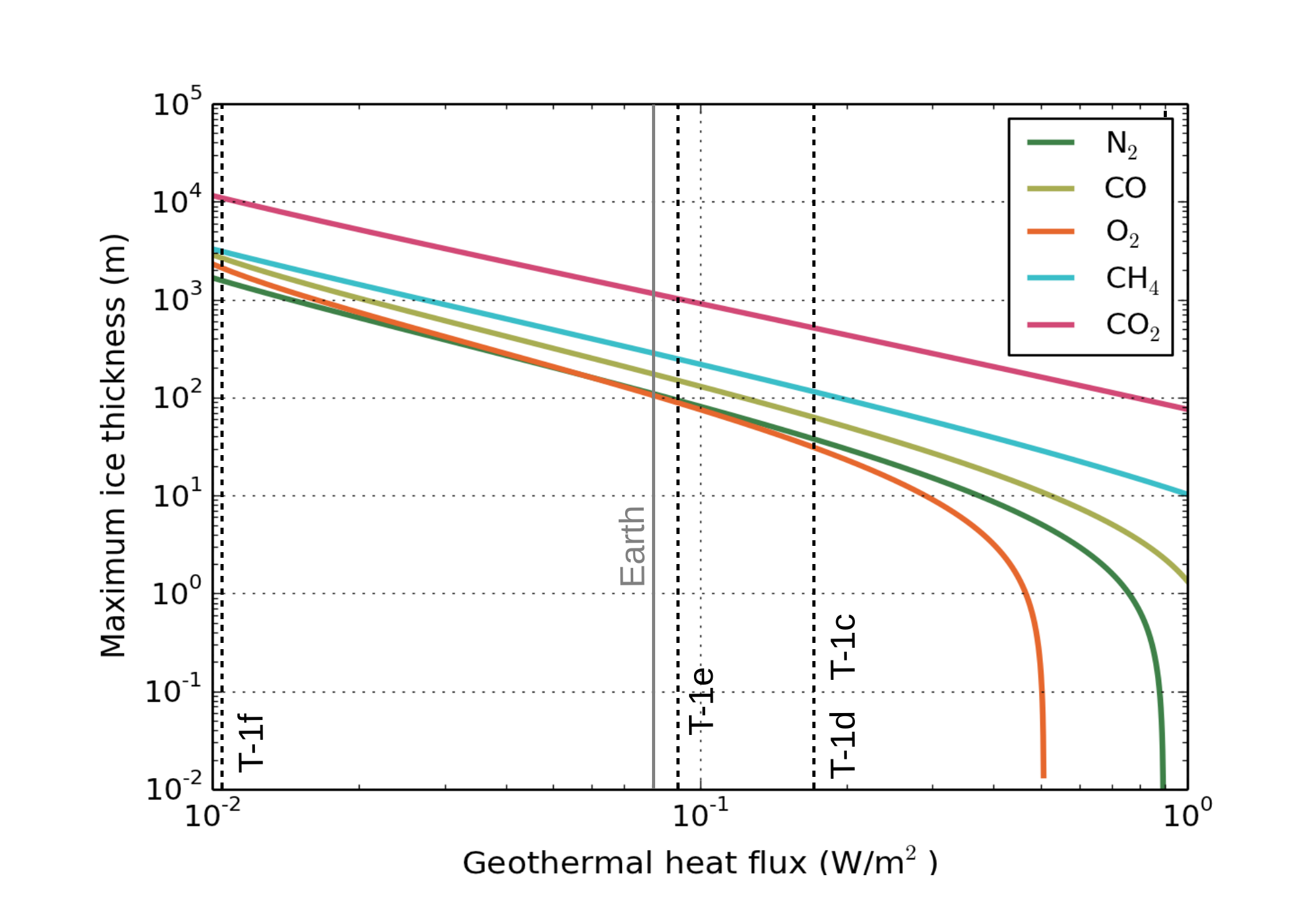}}
\caption{Maximum nightside thickness of various types of ice (N$_2$, CO, O$_2$, CH$_4$ and CO$_2$) - 
assuming that it is limited by basal melting - as a function of the geothermal heat flux. 
It is assumed here that the entire atmosphere has collapsed at the cold points of the planet and 
that the surface temperature at the top of the glacier is controlled by the geothermal heat flux. 
As a reference, vertical dashed lines indicate the average surface tidal heat flux on TRAPPIST-1 planets 
derived from Table~\ref{table_flux}. We also added (vertical solid gray line) the average geothermal heat flux on Earth, 
to give the reader a rough sense of the amplitude of the radiogenic heating on TRAPPIST-1 planets. 
This quantities can be numerically converted in term of global equivalent surface pressure when multiplied by 
a factor $\frac{\rho g}{2}$. The thermodynamical and rheological properties of the ices were taken 
from https://encyclopedia.airliquide.com, http://webbook.nist.gov, \citet{Roder:1978}, \citet{Schm:97}, \citet{Fray2009}, 
\citet{Trowbridge:2016}, and \citet{Umurhan:2017}. Missing rheological data were mimicked on N$_2$.}
\label{ices_thickness}
\end{figure}

At these temperatures that are slightly warmer than those expected at the surface of Pluto \citep{Forget:2017}, 
the viscosity of ices can be rather low. For instance, we estimate from \citealt{Umurhan:2017} (Equation~7) 
that the viscosity of N$_2$ ice at 45~K (resp. 52 and 60~K) should be roughly 1.6$\times$10$^{10}$~Pa~s 
(resp. 8$\times$10$^{8}$ and 7$\times$10$^{7}$~Pa~s). With this condition in mind, it is not clear whether the 
maximum size of glaciers - formed by the accumulation of volatiles - should be controlled by the basal melting 
condition or by the glacial flow. Assessing this question properly would require to compare the efficiency of the 
glacial flow with the rate at which and the position where gaseous N$_2$ would condense on the nightside. 

We assess below the case of the basal melting condition as it gives us an upper limit on the maximum amount of volatile possibly 
trapped on the nightside. When the nightside glaciers start to melt at their base, the ice flow should accelerate and 
expand significantly on the dayside of the planet. Not only basal melting is expected to be a very efficient process to transport ices 
(e.g. N$_2$, CO, O$_2$, etc.) from the nightside to the dayside, but the ices that reach the terminator should get sublimed 
and transport latent heat from the dayside to the nightside. This positive feedback would drive 
the planet into a runaway process, resulting in the formation of a new, global atmosphere.

For any species (we arbitrarily chose N$_2$ here), and for any of the seven TRAPPIST-1 planets, we derive from the 
basal melting condition the following set of 2 equations:
\begin{equation}
\begin{cases}
  T_{\text{base,liq}}~=~T_{\text{ref}}~e^{\frac{\left(\frac{1}{\rho_{\text{liq}}}-\frac{1}{\rho_{\text{ice}}}\right)}
{L_{\text{fus}}}~(g \rho_{\text{ice}} h_{\text{max}}+P_{\text{N}_2}-P_{\text{ref}})}~~\text{,}\\
  h_{\text{max}}~=~\frac{\lambda_{\text{N}_2}~(T_{\text{base,liq}}-T_{\text{night}})}{F_{\text{geo}}}~~\text{,} \\
\end{cases}
\label{n2_equation}
\end{equation}
with $T_{\text{base,liq}}$ the temperature at the bottom of the glacier, $\lambda_{\text{N}_2}$ the conductivity of N$_2$ ice, $\rho_{\text{liq}}$ and $\rho_{\text{ice}}$ the volumetric mass densities of liquid and icy N$_2$, $L_{\text{fus}}$ the latent heat of N$_2$ ice melting, $P_{\text{ref}}$ and $T_{\text{ref}}$ the pressure and temperature of the triple point of N$_2$. P$_{\text{N}_2}$ is the partial pressure of N$_2$ calculated at saturation from the Clausius-Clapeyron relationship (see Fig~\ref{diagram_thermo}) at the surface nightside temperature T$_\text{night}$.

This set of 2 equations corresponds respectively to: 
\begin{enumerate}
    \item the solid/liquid thermodynamical equilibrium at the base of the glacier. Note that the pressure at the bottom of the glacier is controlled by the weight of the glacier (and marginally, by the atmospheric pressure).
    \item the geothermal gradient. It is assumed that the temperature inside the glacier rises linearly with depth, with a lapse rate fixed by the internal heat flux (conductive regime).
\end{enumerate}

This set of equations can be solved explicitly after several variable changes and using the Lambert W function, defined as the solution of X~=~e$^\text{X}$, as done in \citet{Turb:17epsl}.

We calculated and plotted in Figure~\ref{ices_thickness} the nightside maximum thickness as a function of geothermal heat flux for various ices. For geothermal heat flux ranging from 50 to 500~mW~m$^{-2}$, the maximum thicknesses range:
\begin{enumerate}
\item from 200 to 5 meters (Global Equivalent Pressure - GEP - from 15 to 0.2~bar) for N$_2$
\item from 300 to 10~m (GEP from 15 to 0.5~bar) for CO
\item from 200 to 0~m (GEP from 10 to 0~bar) for O$_2$. O$_2$ has the lowest triple point temperature (see Fig~\ref{diagram_thermo}).
\end{enumerate}
Note that these values are of the same order of magnitude than in the atmosphere of Venus ($\sim$~3~bars of N$_2$), Earth (0.78~bar of N$_2$; 0.21~bar of O$_2$) and Titan ($\sim$~1.5~bar of N$_2$), the only Solar System rocky bodies that were able to sustain a thick, global atmosphere.

\begin{enumerate}
\item[$\bullet$] For geothermal heat flux roughly lower than $\sim$~5$\times$10$^2$~mW~m$^{-2}$, 
there is a strong hysteresis on the initial state (volatiles in the atmosphere; or volatiles condensed at the cold points). 
Planets that initially lost their atmosphere could stably accumulate 
quantities of N$_2$/O$_2$/CO up to the equivalent of few bars, 
in condensed form on the surface of their nightside. 
If somehow this scenario occured (through massive accretion or outgassing), 
the volatiles could not be retained on the nightside. The planet would suddenly 
sublime the entire volatile content of N$_2$, CO, O$_2$, CH$_4$, etc. 
forming a brand new, global atmosphere. 

Extreme events such as large meteoritic impact events could also have the potential to destabilize the volatiles that 
have been trapped in condensed form on the surface of the nightside.
Only very eccentric bodies orbiting in the TRAPPIST-1 system could hit the planets near the anti-substellar 
point and potentially sublime the volatiles that should be preferentially trapped there. In the Solar System, 
it has for instance been proposed that the observed distribution of impact craters on Mercury could be explained 
by a large number of very eccentric bodies that would have hit Mercury near the substellar and anti-substellar regions, 
while the planet was in synchronous rotation \citep{Wieczorek:2013}. Note that, there should be generally a large proportion 
of high-eccentricity bodies in the vicinity of the star, favoring subsequently impact events in the anti-substellar region.

\item[$\bullet$] For geothermal heat flux roughly higher than $\sim$~5$\times$10$^2$~mW~m$^{-2}$, we find that the planets could easily form an atmosphere even with very low amount of volatiles. In fact, we start to reach a regime here where the geothermal heat flux itself could significantly contribute to limit the atmospheric collapse as discussed in the previous section. Note that, at such high geothermal heat flux, heat could - and should - be transported by convection; this would significantly alter the calculations made here.
\end{enumerate}

As shown in the previous section, eccentricities of TRAPPIST-1 planets are expected to vary with time, and tidal dissipation and surface heating with it, on timescales $\sim$~1~Earth year (see \citet{Luger2017} - their Supplementary Figure 6). Peaks of tidal surface heating could trigger the destabilization of volatiles trapped on the nightside, although we would expect a delay and smoothing depending on where the tidal dissipation occurs. Actually, the heat itself could alter the internal structure. Detailed calculations of time-dependant tidal-induced surface heat flux (and implications) could be assessed in future studies.

As previously suggested in \citet{Turb:16}, the large-scale gravitational anomalies on tidally locked planets could be aligned with the star-planet axis. This means for instance that it is likely that a large basin (for example impact-induced) could be present at the anti-substellar point of TRAPPIST-1 planets. This could - in the same fashion than nitrogen ice is trapped on Pluto, in Sputnik Planum \citep{Bertrand:2016} - significantly increase the amount of volatiles possibly trapped at the cold point of the planets. Furthermore, the weight of the ices trapped on the nightside could cause the underlying water ice shell to slump, creating by itself (or amplifying the size of) an anti-substellar basin. Such process has recently been proposed as one of the possible scenarios to explain the formation of Sputnik Planitia on Pluto \citep{Hamilton:2016}.

\subsection{Residual atmospheres}

Even though the atmosphere may have collapsed on the cold side of the planet, a residual, thin atmosphere could remain. The volatiles trapped on the nightside should be in fact in thermodynamical equilibrium with a residual atmosphere whose thickness depends on the surface temperature of the nightside, and on the type of volatiles trapped (assuming that the reservoir of volatiles is large enough). 

For a geothermal heat flux of 100~mW~m$^{-2}$ (resp. 200 and 400~mW~m$^{-2}$), 
the temperature of the cold side is $\sim$~36K (resp. 42 and 50~K) and the remnant 
atmosphere could be as thick as $\sim$~0.6~Pa of N$_2$ (resp. 18 and 400~Pa), 
7$\times$10$^{-2}$~Pa of CO (resp. 3 and 110~Pa), and 7$\times$10$^{-3}$~Pa of O$_2$ (resp. 0.6 and 30~Pa). 
For the other volatiles (CH$_4$ and CO$_2$, for example), the thickness (or surface pressure) of a residual atmosphere would be several orders of magnitude lower.

Such residual atmospheres should not be thick enough to significantly increase the global heat redistribution and trigger a N$_2$, CO or O$_2$ runaway process. We remind (from Section~\ref{global_atmosphere}) that the minimum atmospheric pressure required to sustain a global atmosphere is $\sim$~10$^3$~Pa.


Even though detecting a residual atmosphere of N$_2$, CO, O$_2$, etc. might be extremely challenging, as shown above 
such measurements could tell us a lot about 1) the temperature of the nightside and thus the internal heat flux of the planet and 
2) the composition of the nightside reservoir of volatiles.


We also note that volatiles possibly trapped on the nightside of airless close-in planets 
would form a residual atmosphere that would be exposed to various processes of atmospheric escape (mainly stellar-wind sputtering and X/UV-driven hydrodynamic 
escape). This indicates that volatiles trapped on the nightside of geothermally active tidally-locked planets might not be protected from atmospheric escape.



\section{CO$_2$-dominated atmospheres}
\label{co2_section}

All the Solar System terrestrial planets are either airless bodies (e.g. Mercury) or worlds where CO$_2$ is - or was - abundant in the atmosphere (e.g. Venus, Mars) and/or in the subsurface (e.g. Earth). We assume in this section that the four TRAPPIST-1 outer planets possess today large quantities of CO$_2$ either in their atmosphere, on their surface or in their subsurface, and we explore the possible implications.

\subsection{Stability of a CO$_2$-dominated atmosphere}
\label{co2_stab}

\begin{figure*}
\centering
\centerline{\includegraphics[scale=0.8]{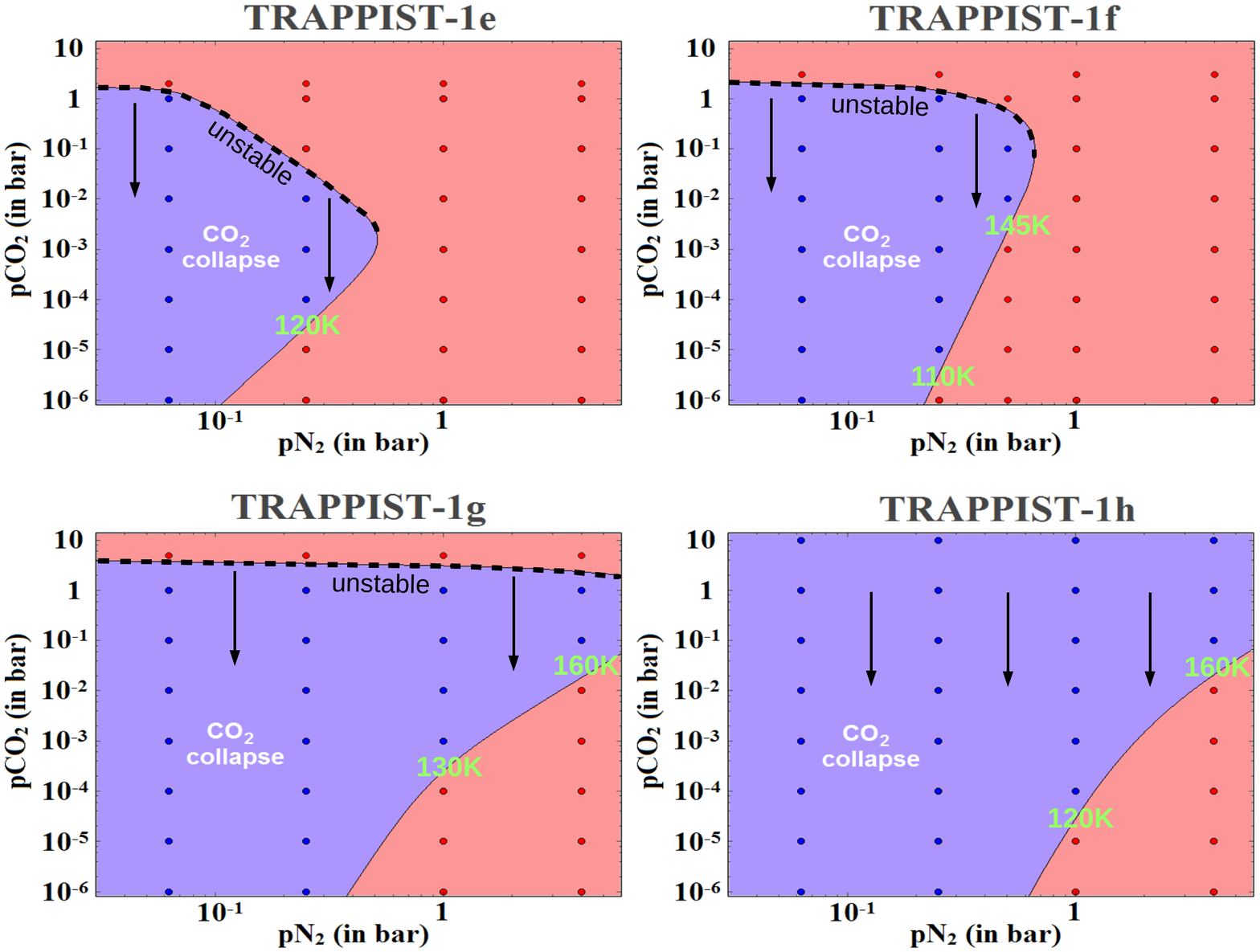}}
\caption{Climate regimes reached as a function of the partial pressures of N$_2$ and CO$_2$. 
For each set of (pN$_2$, pCO$_2$), it is indicated if the atmosphere is stable 
(red) or not (blue) to the atmospheric condensation/collapse of CO$_2$. 
The black arrows indicate how planets that have an unstable atmosphere would 
evolve on this diagram. 
Temperatures (in green) correspond to the rough estimate of the temperature of the cold point, 
at the stable lower boundary (blue is up; red is down). Simulations were performed assuming a surface 
albedo of 0.2 (corresponding both to a water ice surface around TRAPPIST-1, or a rocky surface). 
Water vapor is not included in these simulations. On TRAPPIST-1e, the inclusion of water vapor 
might substantially increase the temperature of the cold points through transport of latent heat from substellar to anti-substellar regions. 
On colder planets, the effect should be marginal.}
\label{n2_co2_collapse}
\end{figure*}

CO$_2$ is much more condensable than any other species discussed in the previous section, as illustrated in Figure~\ref{diagram_thermo}. On synchronously rotating planets, the nightside surface temperature can be extremely low, leading to the condensation of gaseous CO$_2$ on the surface. This could potentially prevent TRAPPIST-1 planets from building up thick CO$_2$ atmospheres.

To test this idea, we performed 130 3D climate numerical simulations of the four TRAPPIST-1 outer planets (surface albedo fixed to 0.2) for atmospheres made of various mixtures of N$_2$ and CO$_2$. In the same vein as \cite{Turb:17epsl}, we find that depending on the partial pressure of background gas (N$_2$, here) and on the partial pressure of CO$_2$, the gaseous CO$_2$ might condense or not, as shown on Figure~\ref{n2_co2_collapse}. The shape of the diagrams is controlled by various physical processes:
\begin{enumerate}
\item The higher the background gas content is, the more efficient the heat redistribution is. This tends to increase the temperature of the cold points and limit the CO$_2$ condensation. High background gas content also favor the pressure broadening of CO$_2$ absorption lines, which increases the greenhouse effect of the atmosphere.
\item The higher the CO$_2$ content is, the higher its greenhouse effect is, but the higher its condensation temperature is. These two processes are competing each other, as illustrated in \citealt{Soto:2015} (in their Figure 1).
\end{enumerate}

Figure~\ref{n2_co2_collapse} shows in fact a bistability in the CO$_2$ atmospheric content. If the planet initially starts with a thick CO$_2$ atmosphere (e.g. 10~bars), the greenhouse effect and the heat redistribution are efficient enough for such atmosphere to be stable (red color). Conversely, if the planet initially starts with a low CO$_2$ atmospheric content or no CO$_2$ at all and progressively accumulates somehow additional CO$_2$ in the atmosphere (e.g. by volcanic outgassing), all the extra CO$_2$ should keep condensing on the nightside (blue color). The planet would thus be permanently locked with a cold, thin CO$_2$ atmosphere.

We can for instance see in Figure~\ref{n2_co2_collapse} that a background atmosphere of $\sim$~100~millibar of N$_2$ is not sufficient to build up a CO$_2$-rich atmosphere from scratch - on any of the four outer planets - due to the nightside surface condensation of CO$_2$. We can also see that TRAPPIST-1h is unable to sustain a dense CO$_2$ atmosphere ($>100$~mbar) even with several bars of N$_2$. For TRAPPIST-1e, f and g, if the initial CO$_2$ content is - for a given amount of background gas - below the "unstable" dotted line, then the planets are unable to build up CO$_2$-rich atmospheres (blue color). However, if the same planets start with an initial CO$_2$ content higher than this limit, CO$_2$ thick atmospheres are found stable (red color).

\subsection{The fate of surface condensed CO$_2$}

What happens when CO$_2$ starts to condense on the nightside? As shown in \cite{Turb:17epsl}, there should be two processes that control the maximum amount of CO$_2$ possibly trapped on the cold side of the planets: 1) CO$_2$ ice flow from the nightside to regions of sublimation; and 2) gravitational burial of CO$_2$ ice beneath water ice cover due to its higher density.

\subsubsection{Glacial flow}

There are in fact two distinct processes that could limit the growth of CO$_2$ ice glaciers:
\begin{enumerate}
    \item The gravity pushes the glaciers to flow from the nightside to the dayside where CO$_2$ ice can be sublimated. This limit depends mostly on the gravity of the planet and the rheological properties of CO$_2$ ice (e.g., viscosity).
    \item The internal heat flux of the planet causes the basal melting of the CO$_2$ ice glaciers. In such conditions, glaciers would slip and flow to the dayside where they would sublimate. This limit depends mostly on the geothermal heat flux of the planet and the thermodynamical properties of CO$_2$ ice (e.g., thermal conductivity).
\end{enumerate}

It has in fact been shown in similar conditions \citep{Turb:17epsl}, 
owing to the low conductivity of CO$_2$ ice ($\lambda_{\text{CO}_2}$~$\sim$~0.5~W$~$m$^{-1}~$K$^{-1}$; \citealt{Schm:97} -- Part I, 
Thermal Conductivity of ices, Figure~4), that it is mostly the basal melting that controls the maximum size of a CO$_2$ ice glacier. 
Using nightside temperatures from GCM simulations (roughly indicated in Figure~\ref{n2_co2_collapse}, in green), we solve a set of 2 
equations similar to Equations~\ref{n2_equation} 
(the only difference being the effect of the partial pressure of the background gas, N$_2$ here) 
to derive the maximum thickness of nightside CO$_2$ ice deposits. These equations are: 
1) the solid/liquid thermodynamical equilibrium at the base of the glacier; and 2) the linear relationship between top and bottom glacier 
temperatures, assuming a fixed lapse rate (conductive regime) forced by the geothermal heat flux.

For 1~bar of background gas (N$_2$), TRAPPIST-1e and f should be protected from CO$_2$ atmospheric collapse. However, for TRAPPIST-1g (resp. h), CO$_2$ could collapse, and as much as 900~/~200~/~80~m of CO$_2$ (resp. 1000~/~250~/~100~m) could be trapped on the nightside for geothermal heat fluxes of 50~/~200~/~500~mW~m$^{-2}$. This corresponds roughly to Global Equivalent Pressure of 45~/~10~/~4~bars (resp. 50~/~12~/~5~bars) of CO$_2$ that could be trapped.

We note that these quantities are of the same order of magnitude than the amount of CO$_2$ outgassed in the Venusian atmosphere ($\sim$~90~bars), or the amount of CO$_2$ contained in the Earth's surface, mostly in the form of carbonate rocks on the continents ($\sim$~10$^2$~bars; \citealt{Walker:1985}).

\subsubsection{When CO$_2$ ice caps are full}

\begin{figure}
\centering
\centerline{\includegraphics[scale=0.2]{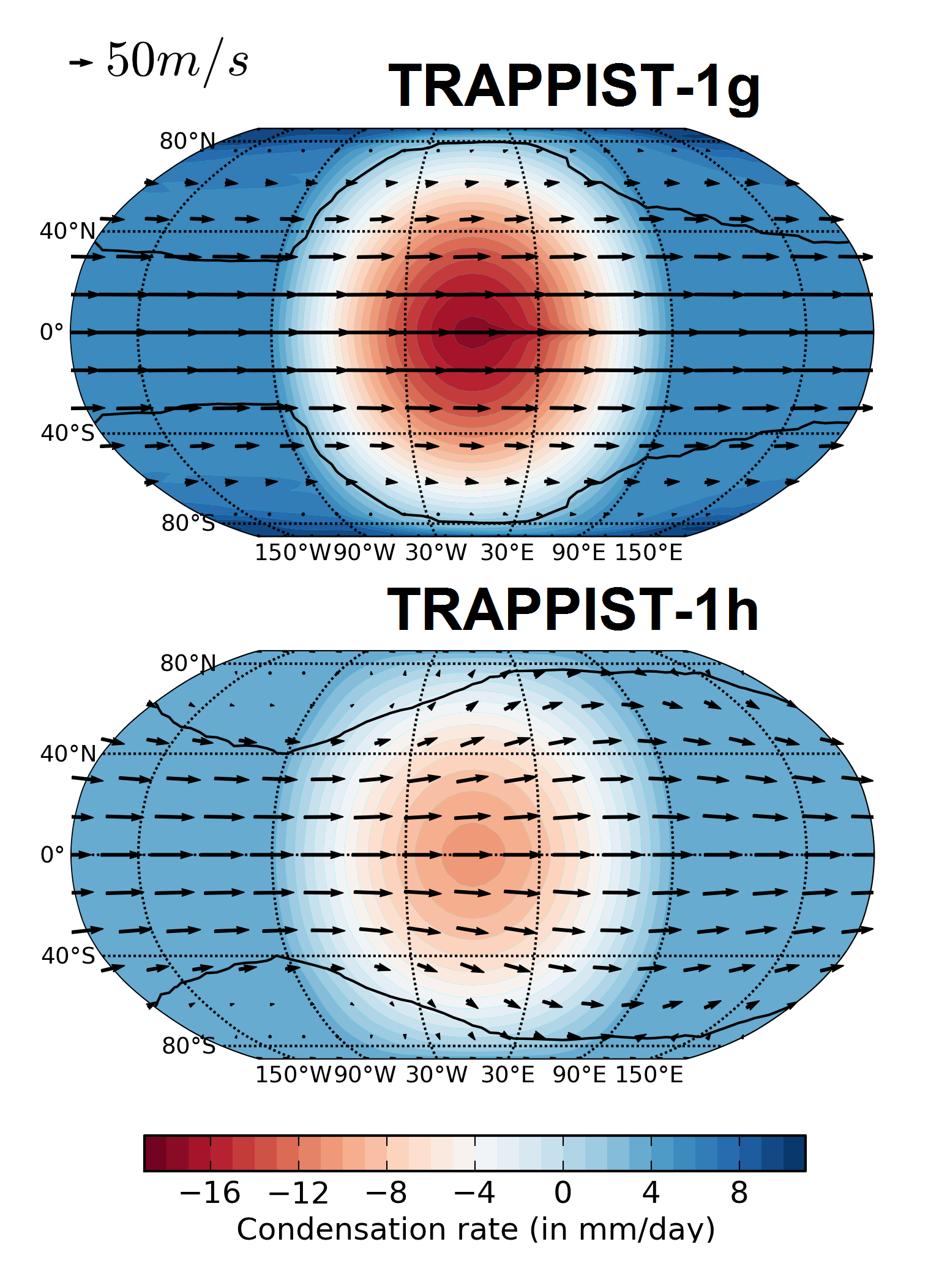}}
\caption{Maps of CO$_2$ ice condensation(+)/sublimation(-) mean day rates (averaged over 50 Earth days) for TRAPPIST-1g and h. 
Wind vectors at 5~km are presented as black arrows (see the 50~m~s$^{-1}$ arrow for the normalization). 
The black line contour indicates the horizontal extent of the CO$_2$ ice clouds (at the 1~g/m$^{-2}$ level).
It is assumed here that the planets are endowed with a pure CO$_2$ atmosphere and have 
with a surface that is entirely covered with CO$_2$ ice. CO$_2$ ice albedo is arbitrarily fixed at 0.5. 
We remind that the radiative effect of CO$_2$ ice clouds is not included here.}
\label{co2_full_cover}
\end{figure}

When the CO$_2$ nightside ice cap becomes "full" (e.g. when CO$_2$ ice starts to convert into liquid), 
all the extra CO$_2$ ice (or liquid) that reaches the irradiated side sublimates/vaporizes into the atmosphere. The extra (now gaseous) CO$_2$ increases the greenhouse effect of the atmosphere. It tends to warm the nightside and thus strengthens the CO$_2$ ice glacial flow, leading to even more CO$_2$ ice sublimation. Depending on the level of irradiation, the planet either finds an equilibrium (with stable CO$_2$ ice/liquid deposits) or enters into a CO$_2$ runaway greenhouse. This scenario has previously been explored for H$_2$O-covered planets \citep{Leco:13} and is extended here to the case of CO$_2$.

It has in fact previously been shown (\citealt{Turb:17epsl}; Appendix C) that CO$_2$ ice caps when full should be unstable on planets:
\begin{enumerate}
 \item that have a low enough geothermal heat flux (typically lower than $\sim$~1~W~m$^{-2}$ for the present study)
 \item and that absorb irradiation fluxes equal or larger than TRAPPIST-1f
\end{enumerate}
CO$_2$ ice caps are expected to be entirely injected in the atmosphere.

To test this idea, we performed 3D GCM simulations of the four TRAPPIST-1 outer planets (efgh) 
endowed with a pure CO$_2$ atmosphere 
where we artificially entirely covered the surface of the planets with CO$_2$ ice 
(with a CO$_2$ ice cover that is large enough that CO$_2$ ice is always present everywhere on the surface). 
CO$_2$ ice albedo is arbitrarily fixed at 0.5. For TRAPPIST-1e and f, we find that no equilibrium is possible. 
The planets cannot maintain surface CO$_2$ ice on their dayside and should always end up in a CO$_2$ runaway greenhouse. 
No equilibrium is possible until 1) all the CO$_2$ ice/liquid content has been sublimed/vaporized, or 2) the CO$_2$ gas content 
becomes so large that CO$_2$ greenhouse effect starts to saturate (whereas the CO$_2$ condensation temperature increases), 
as discussed in \citet{vonparis:2013}.

The scenario is however different for planets that are substantially less irradiated: TRAPPIST-1g and h. 
For these planets, 3D GCM simulations (see Figure~\ref{co2_full_cover}) indicate that an equilibrium where CO$_2$ ice and gaseous coexist is possible. 
In this configuration, the surface temperatures are roughly constant over the planet. For a pure CO$_2$ atmosphere, 
we find an equilibrium at 150$\pm$1~mbar and 174$\pm$0.1K for TRAPPIST-1g (resp. at 4$\pm$0.1~millibar and 145$\pm$0.1K for TRAPPIST-1h). 
The dayside intense CO$_2$ ice sublimation is offset by the nightside condensation, as illustrated in Figure~\ref{co2_full_cover}. For TRAPPIST-1g, approximately 
6~m of CO$_2$ per Earth year (resp. 4~m per Earth year for TRAPPIST-1h) is expected to get sublimed near the substellar point.

This tells us that if TRAPPIST-1g somehow progressively accumulates enough CO$_2$ on its nightside so that it starts to spill on its dayside and get sublimed, the planet should not be able to accumulate enough CO$_2$ in the atmosphere to reach the warm state depicted in Figure~\ref{n2_co2_collapse}. Instead, the planet should be permanently trapped in a cold state, with CO$_2$ ice covering potentially as much as the entire surface of the planet.

Conversely, note that - for TRAPPIST-1g only - if the planet initially starts with a large content of gaseous CO$_2$ (so that it lies above the "unstable" lines, in Figure~\ref{n2_co2_collapse}), then CO$_2$ ice/liquid deposits are unstable.




\subsubsection{CO$_2$ ice gravitational stability}

We assume here that some CO$_2$ has condensed on the nightside of TRAPPIST-1 outer planets, above the water ice shell. CO$_2$ ice is 1.6 times denser than water ice. This difference of density between the two types of ice (CO$_2$ above, H$_2$O below) should trigger an instability of Rayleigh-Taylor \citep{Turb:17epsl} that forces CO$_2$ ice to sink below the water ice cover. At first order, and assuming that both layers of CO$_2$ and H$_2$O ices are isoviscous (i.e. have a fixed viscosity), the density contrast should initiate Rayleigh-Taylor instabilities at a timescale $\tau_{\text{R-T}}$ given by \citep{Turc:01}:
\begin{equation}
\tau_{\text{R-T}}=\frac{13~\eta}{\Delta\rho g b},
\label{eq_turcotte}
\end{equation}
with $\eta$ the viscosity of the more viscous layer, $\Delta\rho$ the density contrast between the two layers and $b$ the characteristic size of the domain. We discuss in the following paragraph how we estimate the different terms of this equation.

Depending on whether CO$_2$ is liquid or solid at the interface with the H$_2$O layer, the density contrast $\Delta\rho$ would range between 240 and 570~kg~m$^{-3}$. Depending on the planet and the background gas content, the surface temperature at the cold point is expected to range between 120~K and 160~K (see Figure~\ref{n2_co2_collapse}). 
However, the basal temperature should rapidly increase with the glacier thickness, given the low conductivity of CO$_2$ \citep{Schm:97}. Assuming that the CO$_2$ glaciers are nearly full, they should have a thickness $\sim$~10$^2$~m and the basal temperature could be as high as $\sim$~218K (temperature at the CO$_2$ liquid/solid equilibrium for a pressure of 1.5~MPa). 
For a stress at the interface between the two layers of the order of 1~MPa and a temperature $\sim$~218~K, the viscosity of the CO$_2$ ice layer is estimated $\sim$~10$^{12}$ Pa~s, based on available experimental data \citep{Durh:99}. 
At the same temperature and stress conditions, water ice has a viscosity $\le$~10$^{16}$~Pa~s, for grain size lower than 1~mm, based on experimental data \citep{Durh:01,Durh:01AREPS,Gold:01}. 
The water ice layer should thus be the layer controlling the Rayleigh-Taylor timescale. Assuming that a thickness of 10$^2$~m of CO$_2$ deposit is representative of the characteristics size of the domain, the R-T timescale $\tau_{\text{R-T}}$ is $\le$~10$^4$~Earth years, which is geologically short. 

Once gravitationally destabilized, the CO$_2$ ice deposit would sink at the base of the water ice shell 
at a rate that is determined mostly by the viscosity of water ice 
and the size of the CO$_2$ ice diapir (e.g. the domed CO$_2$ ice formation piercing the overlying water ice shell). The time required for a CO$_2$ ice 
diapir to cross the water ice layer can be estimated using the Stokes 
velocity, the terminal velocity of a sphere falling through 
a constant viscosity medium \citep{Ziet:07}: 
\begin{equation}
U_s~=~\frac{2}{9}~\Delta\rho g~(r^2 / \eta)
\label{eq_stokes}
\end{equation}
For a diapir radius $r$ of 100~m (comparable to the thickness of the CO$_2$ deposit) 
and a conservative  value for water ice  viscosity of 10$^{15}$-10$^{16}$ Pa~s, this leads to a velocity of 0.04-0.4~m per Earth year. 
As temperature increases as a function of depth ($\sim$~2~F$_{\text{geo}}$~K~m$^{-1}$), the viscosity of water ice 
is expected to decrease with depth, resulting in an acceleration of the diapir fall. 
A 100-m diapir of CO$_2$ ice would thus not need more than $\sim$10$^4$~Earth years to reach the bottom 
of a 1.5-km thick water ice layer, which is the expected depth of a subglacial ocean for a geothermal heat flux $\sim$~0.1~W~m$^{-2}$. 

These two calculations (Rayleigh-Taylor and diapir fall timescales) tell us that 
the lifetime of surface CO$_2$ ice on TRAPPIST-1 planets should be geologically short. 
In particular, it should be short compared to:

\subsubsection*{the volcanic CO$_2$ outgassing timescale}
The present-day Earth CO$_2$ volcanic outgassing rate is 60~bars/Gy \citep{Brantley:1995, Jarrard:2003}. 
It takes roughly 10$^6$ Earth years to outgass a $\sim$60~cm Global Equivalent Layer (GEL) of CO$_2$ ice 
(equivalent to a 10$^2$m-thick nightside CO$_2$ ice cap with a radius of 10$^3$~km).

\subsubsection*{the CO$_2$ ice flow and sublimation timescale}
We assume that a 10$^2$m-thick, 10$^3$km-radius nightside CO$_2$ ice cap is in dynamic equilibrium with the atmosphere. 
This means that the CO$_2$ ice flow - controlled here 
by the rheological properties of CO$_2$ ice - has reached a constant, positive rate. 
This also means that the integrated CO$_2$ ice sublimation rate at the edges of the glacier 
is equal to the total gaseous CO$_2$ condensation rate on the ice cap.

We model the steady state flow of the CO$_2$ ice cap using equations~1-4 from \citet{Meno:13}.
We take the flow rate constants of CO$_2$ ice from \citet{Nye:2000}, derived from the measurements of \citet{Durh:99}. 
We chose the rheological properties of CO$_2$ ice for a creep exponent n=2, and at 218K. This is the maximum temperature expected 
at the bottom of the CO$_2$ ice glacier, before basal melting occurs. These are conservative assumptions in the sense that 
these are the parameters (creep law and temperature) that maximize the velocity of the CO$_2$ ice flow.
With these assumptions, we estimate that it takes at least 10$^8$ Earth years to recycle the entire CO$_2$ ice cap.

\subsubsection{The fate of buried CO$_2$}

CO$_2$ ice is expected to completely melt and equilibrate thermally with the surrounding H$_2$O media when stabilized at the bottom of the water ice shell. 
The temperature and pressure conditions at the bottom of the water ice layer depend on its thickness and on the geothermal flow. For geothermal heat flux lower than $\sim$~0.4~W~m$^{-2}$, the melting of water ice would be reached for depth larger than $\sim$~4$\times$10$^2$~m, and pressure of $\sim$~3.5~MPa, corresponding to the saturation vapor pressure of CO$_2$ at $\sim$~273~K \citep{Lide:04}. 
Destabilizing the liquid CO$_2$ would therefore require a geothermal heat flux higher than 0.4~W~m$^{-2}$. At such large geothermal heat flux, CO$_2$ ice (or liquid) should indeed get sublimed (or vaporized) within the water ice shell.

However, for geothermal heat flux lower than 0.4~W~m$^{-2}$, CO$_2$ ices/liquids should be stable during their fall. Even if the density of liquid CO$_2$ decreases with increasing temperature as it equilibrates with the surrounding water ice media, it remains always denser than water ice \citep{Span:96}, and therefore should always accumulate at the bottom of the ice shell. At T~=~273~K and pressure between 3.5 and 28~MPa (subglacial pressures estimated for geothermal heat flux between 400 and 50~mW~$^{-2}$), liquid CO$_2$ has a density very close to that of 
liquid water (928 and 1048 kg~m$^{-3}$, respectively, using the equation of state of \citealt{Span:96}), so that CO$_2$ should coexist with H$_2$O at the ice-water interface. 

From this point, two processes are expected to occur and compete with each other. Firstly, part of the CO$_2$ should dissolve in the liquid water. The total amount of CO$_2$ that could be dissolved in the water layer would depend on the volume (thickness) of the water layer. 

Secondly, pressure and temperature conditions expected at the bottom of the ice layer are in the stability field of CO$_2$ clathrate hydrate \citep{Sloa:98,Long:05}, therefore CO$_2$ should rapidly interact with H$_2$O molecules to form clathrate hydrate. 
Clathrate hydrates are non-stoichiometric compounds consisting of hydrogen-bonded H$_2$O molecules forming cage-like structures in which guest gas molecules, such as CO$_2$, can be trapped \citep{Sloa:98}. Once formed, these clathrates are very stable and can be dissociated only if the temperature is raised about 5-10~K above the melting point of water ice. The storage of CO$_2$ in the form of 
clathrate should be particularly efficient as liquid CO$_2$ and liquid water coexist. 
As CO$_2$ clathrate hydrates have a density of about 1150 kg~m$^{-3}$ (assuming full cage occupancy, \citealt{Sloa:98}), they would rapidly sink at the bottom of the water liquid layer, ensuring an almost complete clathration of CO$_2$. Note that we expect some of the CO$_2$ to get dissolved in the liquid water during the clathrate sinking. The relative proportion of CO$_2$ trapped in the form of clathrate hydrate or dissolved in the water layer 
would depend on the volume of CO$_2$ that is buried at the base of the ice shell and on the volume (thickness) of the water layer.

In summary, as long as the geothermal heat flux is lower than $\sim$~0.4~W~m$^{-2}$, the water ice shell should exceed several hundreds of meters, and CO$_2$ should remain sequestered either in the form of CO$_2$ clathrate hydrates or dissolved in the subglacial liquid water ocean. Release of gaseous CO$_2$ in the atmosphere may occur in particular following local increase of geothermal heat flux resulting in a significant thinning and breaking-up of the water ice shell. The total amount of CO$_2$ that can be stored in the H$_2$O layer (by any of the two processes discussed above) depends on the total abundance of H$_2$O of the planet as well as the CO$_2$/H$_2$O ratio. 
Evaluating the maximum amount of CO$_2$ that can be trapped underneath the water 
ice cover require a detailed description of the H$_2$O layer structure as well 
as thermodynamic models predicting the partitionning of CO$_2$ between the different phases.



\section{CH$_4$-dominated worlds}
\label{reduced_section}

\begin{figure*}
\centering
\centerline{\includegraphics[scale=0.2]{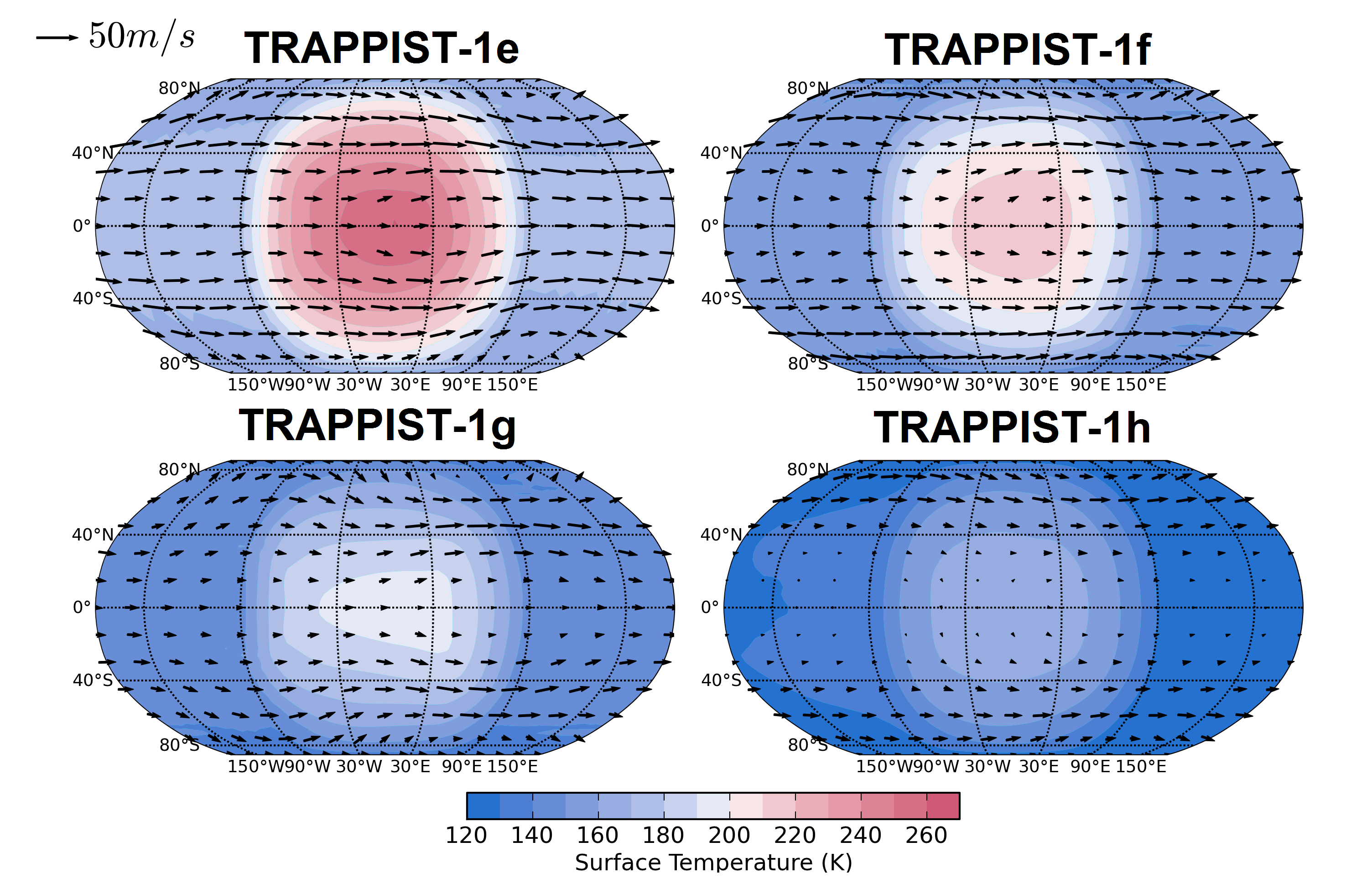}}
\caption{Maps of surface temperatures (averaged over 50 Earth days) for TRAPPIST-1e, f, g and h, assuming planets 
endowed with a 1~bar N$_2$-dominated atmosphere composed of 10$\%$ of CH$_4$. 
Wind vectors at 5~km are presented as black arrows (see the 50~m~s$^{-1}$ arrow for the normalization). 
Surface albedo is arbitrarily fixed at 0.2.}
\label{n2_ch4_3D}
\end{figure*}

\begin{figure*}
\centering
\centerline{\includegraphics[scale=0.95]{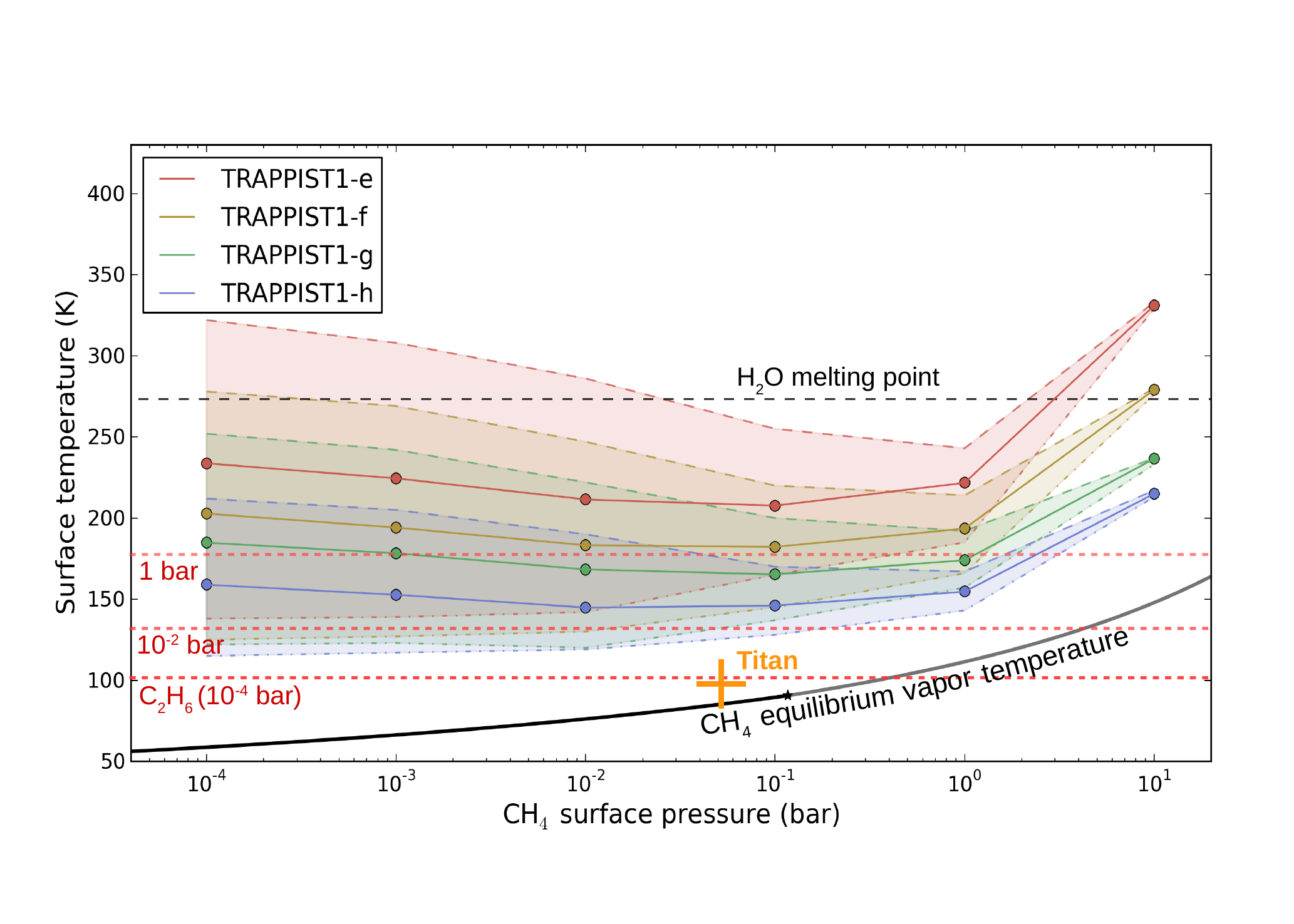}}
\caption{Mean surface temperatures of TRAPPIST-1 outer planets, assuming atmospheres made of N$_2$ and CH$_4$ only. Each simulation was performed with the 3D LMD Generic Global Climate Model, for 1 bar of N$_2$, and various CH$_4$ partial pressures (from 10~Pa to 10~bars). The four regions filled in colors show the range of surface temperatures reached by the four outer planets (red, yellow, green and blue for TRAPPIST-1e, f, g and h, respectively). Solid, dashed and dash-dotted lines depict the mean, maximum and minimum surface temperatures reached in the simulations, respectively. Note that the highest surface temperature for TRAPPIST-1f, g and h is almost always lower than 273~K (e.g. the melting point of water). The black line indicates the CH$_4$ equilibrium vapor pressure. Dashed red lines show the equilibrium vapor temperature of ethane (C$_2$H$_6$) for various partial pressures (10$^{-2}$, 10$^{-4}$; and 1~bar). The melting point of water is indicated by the black horizontal dashed line. As a reminder, equilibrium temperatures of TRAPPIST-1e, f, g and h planets are (for a Titan-like bond albedo of 0.3) respectively 230, 198, 182 and 157~K.}
\label{n2_ch4_tsurf}
\end{figure*}

The only Solar System terrestrial-size object that possesses a thick atmosphere that deviates from the one discussed as far is Titan. Titan (0.012~$\Searth$, 0.4~$\Rearth$) has a 1.5~bars thick N$_2$-dominated atmosphere, with as much as 5~$\%$ of methane near the surface \citep{Niemann:2005}.
We explore in this section the possibility that TRAPPIST-1 outer planets could be hydrocarbon-rich worlds, and the possible implications.

\subsection{Warm Titans}
\label{warm_titans}

What would happen if you suddenly place Titan at the location of each of the seven TRAPPIST-1 planets and how would the planet evolve?
At the equilibrium temperatures of the four TRAPPIST-1 outer planets (e to h), the saturation vapor pressure of CH$_4$ ranges between 10 and 100~bars (resp. between 5$\times$10$^{-2}$ and 5~bars for C$_2$H$_6$). Unlike Titan, we should thus expect all the methane and/or ethane content to be vaporized in the atmosphere.

To check this, we performed haze-free 3D numerical climate simulations 
of N$_2$/CH$_4$ atmospheres, for various CH$_4$ contents, and for the four TRAPPIST-1 outer planets.
Some of these simulations (for a 1~bar N$_2$-dominated atmosphere with 0.1~bar of CH$_4$, similar to Titan) 
are presented in Figure \ref{n2_ch4_3D}.. 
Fig~\ref{n2_ch4_tsurf} shows the mean, maximum and minimum surface temperatures obtained for each of the TRAPPIST-1e, f, g and h outer planets. 
The calculated surface temperature of the planets results from a subtle balance between the radiative cooling of stratospheric 
CH$_4$ and the greenhouse effect of tropospheric CH$_4$. Around a star like TRAPPIST-1, absorption of stellar radiation by CH$_4$ 
is particularly efficient around 0.89, 1.15, 1.35, 1.65, 2.3 and marginally 3.3~$\mu$m bands. Consequently, CH$_4$ absorption warms 
the upper atmosphere and also reduces the short wave irradiation flux that reaches the surface and troposphere, contributing to a cooling 
of the planetary surface. For example, approximately 40$\%$ of the incoming stellar radiation 
is able to reach the surface in the simulations shown in Figure~\ref{n2_ch4_3D}.

Despite the anti-greenhouse effect of CH$_4$ that tends to cool the surface temperature, it would be extremely hard for TRAPPIST-1 planets to sustain surface liquid/icy CH$_4$. This is illustrated in Fig~\ref{n2_ch4_tsurf} with the comparison between the saturation pressure curve of CH$_4$ and the calculated minimum surface temperatures (on the nightside). Additionally, a partial pressure of at least $\sim$~10$^{-2}$~bar of ethane (C$_2$H$_6$) would be required for the coldest planets to start forming nightside surface lakes or seas of ethane, similar to the ones observed on Titan.

UV flux should lead to the formation of photochemical hazes in CH$_4$-rich planetary atmospheres, 
in the same fashion than on Titan. Such hazes could potentially have an additional powerful anti-greenhouse 
(i.e. radiative cooling) effect, by absorbing and reflecting a significant part of the incoming stellar flux \citep{Lorenz:1997}. 
TRAPPIST-1 outer planets (e, f g and h) should receive a EUV flux ranging between 600 and 3000 times Titan's flux \citep{Wheatley:2017}. 
Photochemical hazes could thus form efficiently and accumulate in the atmosphere. This could potentially cause a catastrophic cooling of 
the planetary surface \citep{Mckay:1991}, and change the aforementioned conclusions on the likeliness for TRAPPIST-1 outer planets to sustain 
surface condensed methane.

Nonetheless, using a 1D photochemical-climate model taking properly into account the microphysics and the radiative effect 
of photochemical hazes \citep{Arney_phd:2016}, it has been shown that the thickness (and thus the opacity) of organic hazes 
should be self-regulated \citep{Arney:2016}. In fact, thick hazes should inhibit methane photolysis, which would in turn drastically 
limit haze production rates. In other words, the rate of methane photolysis should not scale linearly with the incoming UV flux, 
but instead should at some point saturate.
Moreover, organic hazes are much less opaque at the emission wavelengths of cool stars like TRAPPIST-1 
than solar emission ones \citep{Khare:1984,Vinatier:2012,Arney:2017}. This indicates that a large part of the incoming stellar 
flux would reach TRAPPIST-1 planetary surfaces and tropospheres and easily vaporize all the methane in the atmosphere. 
In other words, even when taking into account the radiative effect of hazes, all the TRAPPIST-1 planets should be well 
beyond the CH$_4$ runaway-greenhouse-like limit.
Eventually, it is important to note that, at the high EUV fluxes expected on TRAPPIST-1 planets, CO$_2$ (if present) 
could be also photodissociated into oxygen radicals that should seriously limit the build up of the organic hazes 
\citep{Arney:2017}. In particular, if the atmospheric CO$_2$/CH$_4$ ratio is high, and if the emission of TRAPPIST-1 
is high in the spectral region $\sim$~[120-180]~nm, where the UV cross section of CO$_2$~/~O$_2$ is maximum 
(\citet{Arney:2017}, Figure~2c), then the formation of photochemical hazes could be severely halted.

We remind that a potentially thick O$_2$ atmosphere could have built up abiotically during the early runaway phase \citep{Luger:2015,Bolmont2017} while TRAPPIST-1 was a pre-main-sequence star, playing potentially a strong role here on the haze formation. But above all, the combustion of CH$_4$ (and more generally, of any reduced compound such as NH$_3$ or H$_2$S), following
CH$_4$~+~2O$_2$~$\rightarrow$~CO$_2$~+~2H$_2$O
, should prevent CH$_4$ to substantially build up in a thick O$_2$-rich atmosphere. If the build-up of O$_2$ during the early runaway phase exceeds the total reservoir of CH$_4$, there might not be enough room left for CH$_4$ to accumulate in the atmosphere.

\subsection{Titan-like world lifetime}

Through 1) CH$_4$ and hydrocarbons photodissociation, 2) organic hazes formation and 3) haze sedimentation, the atmospheric CH$_4$ and hydrocarbon content of TRAPPIST-1 planets should deplete rapidly. 
It is for example estimated that it should take roughly 10~My for Titan to remove all the methane (0.07~bar) from the atmosphere \citep{Yung:1984}, and as much as $\sim$ 30~bars could have been destroyed since the beginning of the Solar System. 

Therefore, as much as 600-3000 times more methane (averaged over the surface) could potentially be photolyzed on TRAPPIST-1 outer planets. Over the expected age of the TRAPPIST-1 system (between 5 and 10~Gyr, according to \citealt{Luger2017} and \citealt{Burgasser2017}), at least $\sim$~120~bars of CH$_4$ (Titan's limit, including the gravity correction) and as much as 10$^5$~bars of CH$_4$ (when scaling linearly the CH$_4$ loss with the EUV flux) could have been destroyed by photolysis.

Sustaining continuously a CH$_4$-rich (and NH$_3$-rich, by analogy) atmosphere over TRAPPIST-1 lifetime would 
require an extremely large source of methane. It is in fact widely believed that the CH$_4$ current level on Titan might be somewhat anormal and produced by an episodic replenishment due to destabilization of methane clathrates in Titan's subsurface \citep{Tobie:2006}.

Similarly, large quantities of N$_2$ could be photodissociated, 
forming HCN \citep{Liang:2007,Krasnopolsky:2009,Tian:2011,Krasnopolsky:2014}, and could be lost subsequently in 
longer carbonated chains that could sedimentate on the surface (see next subsection). 
This mechanism could remove efficiently N$_2$ from the atmosphere in the long term.

We acknowledge however that the arguments stated in the previous section (especially the haze negative feedback on CH$_4$ photolysis, as proposed by \citealt{Arney:2016}) could drastically limit the CH$_4$ photolysis rate and relax the constraint on the methane production rate required to sustain a CH$_4$-rich atmosphere.


\subsection{Surface conditions}

Even for large CH$_4$/N$_2$ contents, and even when neglecting the radiative effect of photochemical hazes, TRAPPIST-1f, g and h should be cold enough (see Figure~\ref{n2_ch4_tsurf} and the associated legend) to be covered by a complete layer of water ice. In this case, photolysis of methane would produce organic hazes that should sedimentate and progressively accumulate at the surface in large quantities. On Titan, it is estimated that $\sim$~1~m Global Equivalent Layer (GEL) of heavy hydrocarbons - or tholins - are covering the surface \citep{Lorenz:2008}. This is in fact two orders of magnitude lower than what we would expect from the direct conversion of current CH$_4$ photolysis rate through the age of the Solar System \citep{Lorenz:1996}. Possible solutions to this discrepency are discussed in \citet{Lorenz:2008}. 

Similarly, signatures of long carbonated chains have also been detected on many Kuiper Belt Objects (\citealt{Johnson:2015}; see Table~1 and references therein), including Pluto, Triton, Makemake, Sedna, etc. The New Horizon mission has even directly observed and mapped (during its flyby) dark tholins deposits on Pluto, in Cthulhu Regio \citep{Stern:2015}. 

TRAPPIST-1 planets could thus be covered by a thick surface layer of tholins today. Once the CH$_4$ atmospheric reservoir would be empty, only condensable hydrocarbons and long-carbonated chains (and potentially gaseous N$_2$, leftover from NH$_3$) would remain. In general, because sedimented organic hazes should be rather rich in hydrogen, they should have a density of the same order of magnitude than water ice, and should be in particular more stable than CO$_2$ to gravitational burial.


Assuming that photochemical hazes have a limited radiative effect in the near-infrared \citep{Khare:1984,Vinatier:2012,Arney:2017} 
where the emission of TRAPPIST-1 peaks, 
large quantities of CO$_2$ (added to CH$_4$/N$_2$ and other greenhouse gases) could be 
sufficient to raise the surface temperature of TRAPPIST-1 outer planets above the melting point of water, 
although this need to be tested with coupled photochemical~/~3-D climate model in the future. 
In this case, sedimented organic carbonated chains should not accumulate at the surface but instead 
should get dissolved in the liquid water ocean. This, and the UV shield provided by the photochemical 
hazes \citep{Wolf:2010,Arney:2017} could provide TRAPPIST-1 planets with surface conditions 
favorable for life - as we know it - to emerge and develop.

We discuss more generally in the next section the conditions required for TRAPPIST-1 planets to sustain surface habitability.



\section{The habitability of TRAPPIST-1 planets}
\label{trappist_habitability}

Most of our knowledge on habitability comes from the study of Venus, Mars, and Earth. The system of TRAPPIST-1 displays a fantastic zoology of planets to confront our theories with, and potentially revolutionize all what we know on this domain.

\subsection{The habitability of the inner planets TRAPPIST-1bcd}

The two inner planets of the system (TRAPPIST-1b and c) are likely too hot to sustain global oceans of liquid water \citep{Kopparapu2013,Yang2013,Kopparapu2016}. Nonetheless, they could still be desert worlds with limited surface water \citep{Abe:2011} trapped in nightside niches (e.g. land planets) or at the edge of large scale glaciers near the terminator \citep{Leco:13}.

TRAPPIST-1d (S$_\text{eff}$~$\sim$~1.14~$\Searth$) however is near the inner edge of the Habitable Zone of synchronously-rotating planets \citep{Yang2013,Kopparapu2016}. If TRAPPIST-1d is able somehow to sustain a thick, highly reflective water cloud cover near the substellar region, it could sustain surface liquid water global oceans. Detailed 3D modeling of clouds, and more generally of all the possible parameters that could affect the atmospheric circulation, would be required to assess this possibility.

\subsection{The remarkable potential of TRAPPIST-1e for habitability}

\begin{figure*}
\centering
\centerline{\includegraphics[scale=0.2]{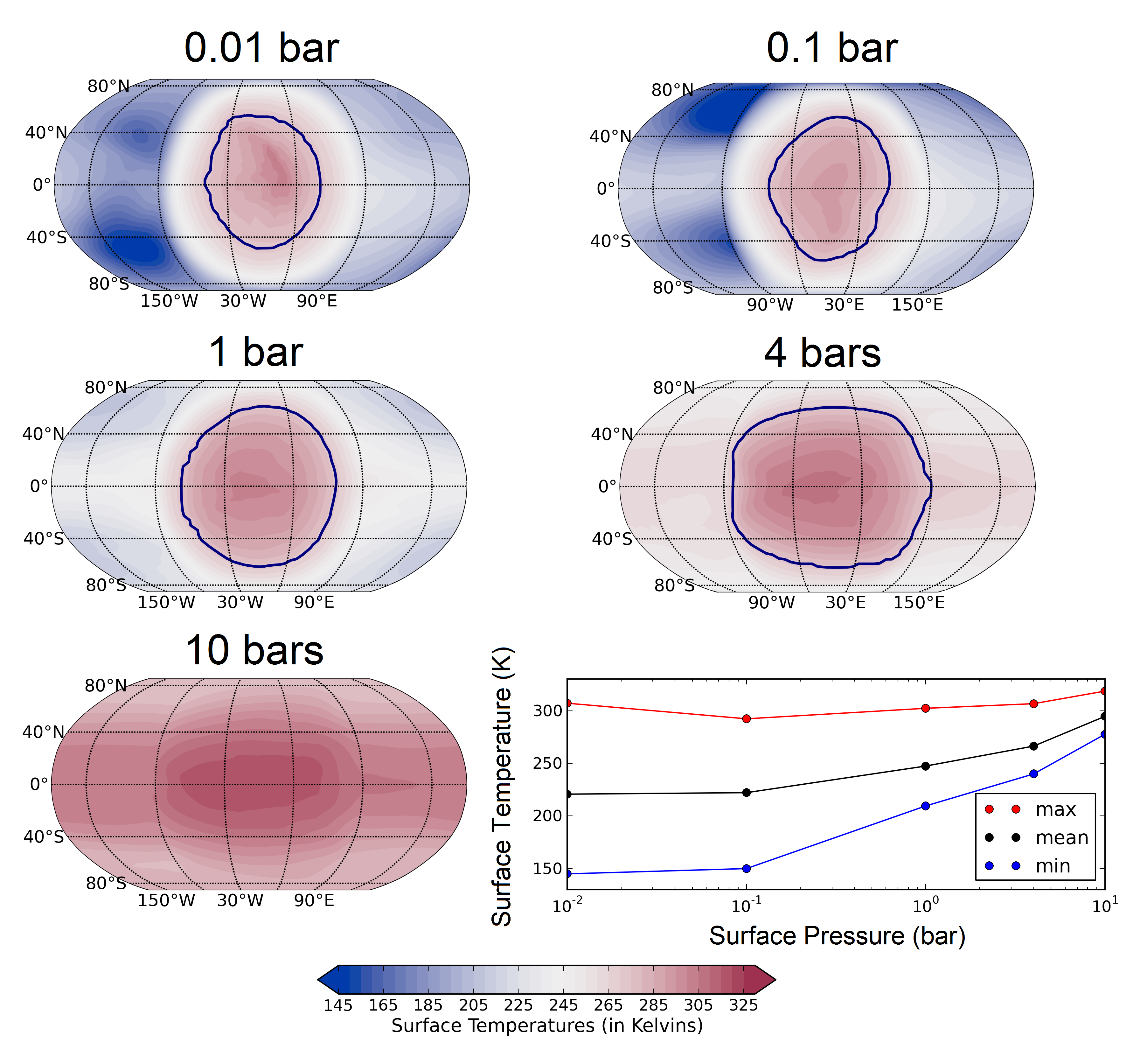}}
\caption{4-years average surface temperature maps of TRAPPIST-1e endowed with atmospheres made of N$_2$ and 376~ppm of CO$_2$, and for various atmospheric pressures (10~mbar, 0.1~bar, 1~bar, 4~bars and 10~bars). Solid line contours correspond to the delimitation between surface liquid water and sea water ice. The Figure in the bottom right panel indicates in blue, black and red the minimum, mean, and maximum surface temperatures, respectively. Note that the planets were assumed to be initially cold (T~=~210~K everywhere) and completely covered by water ice.}
\label{map_T1E}
\end{figure*}

According to our simulations, TRAPPIST-1e is the only planet in its system with the ability to host surface liquid water without the need of greenhouse warming from another gas than H$_2$O. This requires a sufficient H$_2$O reservoir covering the whole surface (i.e. that cannot be fully trapped on the nightside). Thanks to the synchronous rotation, the received stellar flux (F$\sim$904~W~m$^{-2}$) is sufficient to maintain a least a patch of liquid water at the substellar point, even in the absence of a background atmosphere. This configuration is usually known as the eyeball regime \citep{Pierrehumbert2011}.

This situation is similar to that of Proxima Cen b \citep{Anglada16}. This potentially rocky (most probable mass of 1.4$M_\oplus$) planet orbiting the closest star from our Sun receives within uncertainties nearly the same amount of stellar energy (F$\sim$890~W~m$^{-2}$) as TRAPPIST-1e. Two studies with two different GCMs \citep{Turb:16,Boutle:2017} showed that a water-rich and synchronous scenario for Proxima~b generates a substellar surface ocean. The reader is refered to \citealt{Turb:16} (in particular, their Figure~1) for a detailed discussion on the possible climate regimes on TRAPPIST-1e, analogous to Proxima Cen b.

We performed several 3D GCM simulations (see Fig~\ref{map_T1E}), assuming a cold start (T~=~210~K everywhere, full water ice coverage). We find that for any atmosphere, TRAPPIST-1e always ends up with surface liquid water, at least in the substellar region. 
This would hold even with no background atmosphere at all; in this case, the atmosphere would be composed of water vapor. Starting from this point, adding greenhouse gases to the atmosphere would increase the mean surface temperature and increase the size of the patch of liquid water.
We not only confirm here that the case of TRAPPIST-1e is analogous to Proxima Cen b, but we also show that, due to the lowered albedo of water ice around TRAPPIST-1, these conclusions do not depend on the initial state.

In summary, if 1) TRAPPIST-1e is in synchronous rotation and 2) is water-rich, then the planet should have a patch of 
liquid water at its substellar point, whatever its atmosphere (as thin or thick as wanted) and whatever its 
initial state (fully glaciated or not). This result must hold for any arbitrary atmospheric composition, 
unless a tremendous anti-greenhouse effect occurs (e.g. absorption by stratospheric methane, or absorption and reflection by photochemical hazes), 
or unless a tremendous greenhouse effect (by a thick Venus-like atmosphere, for example) raises the mean surface temperature above 
the critical point of water (647K). 
These possibilities will be explored in details in future studies.

If low density estimates of TRAPPIST-1e \citep{Gillon2017,Wang:2017} were to be confirmed in the future, 
indicating that the planet could have retained large quantities of water, 
TRAPPIST-1e would thus become a fantastic candidate for habitability outside our Solar System. More generally, TRAPPIST-1e together with Proxima Cen b highlight a new type of planets that should always sustain surface liquid water, and that are therefore extremely promising for habitability prospects.

\subsection{The habitability of outer planets}

Besides TRAPPIST-1e, the three outer planets of the system (TRAPPIST-1f, g and h) are interesting probes to study habitability outside our Solar System.

\subsubsection{Surface habitability cannot be sustained with background gases only}

From the set of simulations described in Section~\ref{global_atmosphere} and extended to atmospheric pressures as thick as 4~bars, we find that none of the 3 TRAPPIST-1 outer planets (f, g and h) are able to maintain surface liquid water, assuming a background atmosphere - not able to generate a significant greenhouse effect - that would only be made of N$_2$, CO or O$_2$ (with H$_2$O included as a variable species). 

It tells us that TRAPPIST-1f, g and h need to build up greenhouse gases in their atmosphere (e.g. CO$_2$, CH$_4$, NH$_3$, H$_2$, etc.) to sustain surface habitability. We explore this possibility in the next sections.

\subsubsection{Minimum CO$_2$ content required for surface habitability}

Using 3D and 1D (with our one-dimensional cloud-free climate model \citep{Wordsworth:2010aa} that uses 
the same physical package than the 3-D LMD Generic GCM described in Section~\ref{lmd_gcm}) simulations 
of planets endowed with thick CO$_2$-dominated atmospheres, we find that:
\begin{enumerate}
\item Planet f can maintain surface liquid water for CO$_2$-dominated atmosphere thicker than $\sim$~1~bar. Note that a warm solution is possible for lower CO$_2$ atmospheric pressures, although CO$_2$ would condense on the surface, leading to a complete atmospheric collapse.
\item Planet g can maintain surface liquid water for CO$_2$-dominated atmospheres thicker than $\sim$~5~bars.
\item Planet h is not suitable for surface liquid water, whatever the thickness of the CO$_2$ atmosphere is, and even when maximizing the radiative effect of CO$_2$ ice clouds (parameterized following \citealt{Forg:13}). In fact, Figure~\ref{n2_co2_collapse} also tells us that TRAPPIST-1h is unable to build up a CO$_2$ atmosphere, whatever the background gas content. For a 1~bar N$_2$ atmosphere, TRAPPIST-1h should not be able to build up more than few tens of ppm of CO$_2$ in the atmosphere.
\end{enumerate}

Our results are roughly in agreement with the reference papers by \cite{Kopparapu2013} and \cite{Kopparapu2014} on habitability. We note however that the recent paper by \citet{Wolf:2017}, which finds that CO$_2$-dominated atmospheres as thick as 30~bars cannot warm the surface of TRAPPIST-1f and g above the melting point of water, is at odd with our results, and more generally, with the related literature. We believe that the discrepancy comes potentially from the fact that \citet{Wolf:2017} underestimated in his calculations the effect of the lowered albedo of ice~/~snow around cool stars (mean water ice albedo of 0.21 around TRAPPIST-1) due to the shape of its reflectance spectrum, as supported by experimental data \citep{Warren1980,Joshi2012}. This would suppress the runaway glaciation positive feedback invoked by \citet{Wolf:2017}. 

The discrepency could also come from differences in the radiative treatment of CO$_2$-rich atmospheres. We used here the parameterization of CO$_2$ absorption lines following \citet{Wordsworth:2010}, using updated line intensities and positions, and half-width at half-maximum from HITRAN-2012 \citep{Rothman2012}, sublorentzian line shape of \citet{Perrin:1989} and Collision-Induced Absorptions (CIA) of \citet{Gruszka:1997} and \citet{Baranov:2004}. \citet{Wolf:2017} used instead CO$_2$ cross sections from \citet{Wolf:2013}, based on HITRAN-2004, and did not include the effect of Collision Induced Absorptions (CIA) and dimer absorptions despite their importance when modelling thick CO$_2$ atmospheres \citep{Wordsworth2010,TurbetTran:2017}.

More generally, using our 1D~/~3D GCM simulations, we find that the outer edge of the classical Habitable Zone around TRAPPIST-1 (using the TRAPPIST-1 synthetic spectrum, and for an atmosphere of 70~bars of CO$_2$) lies around 306~W~m$^{-2}$ (S$_{\text{eff}}$~=~0.225). This value is slightly higher than the one (302~W~m$^{-2}$; S$_{\text{eff}}$~=~0.221) given in \cite{Kopparapu2013}. 

We explored the effect of 1) gravity, 2) rotation mode, 3) changing the stellar spectrum (from synthetic to blackbody), and 4) including the radiative effect of CO$_2$ ice clouds, and found that their cumulative effect on the limit of the outer edge of the Habitable Zone should not exceed roughly 30~W~m$^{-2}$ in the context of the TRAPPIST-1 exoplanetary system. 

\subsubsection{The case of TRAPPIST-1h}

TRAPPIST-1h is a poor candidate for surface habitability, and here are the following reasons why:

\begin{enumerate}
\item[$\bullet$] As explained in Section~\ref{co2_stab}, TRAPPIST-1h is unable to accumulate a dense CO$_2$ atmosphere (because of surface condensation) that could warm the surface and favor surface habitability.
\item[$\bullet$] As shown in Section~\ref{reduced_section}, even 1) when considering an unlikely scenario where a CH$_4$ thick atmosphere would have been built, and even 2) when neglecting the radiative cooling of photochemical hazes, we find that (see Fig~\ref{n2_ch4_tsurf}) CH$_4$-dominated atmospheres are unable to raise the mean surface temperature of TRAPPIST-1h above $\sim$~160~K for an atmosphere made of 1~bar of N$_2$ and CH$_4$ content lower than 1~bar. This result is mostly due to the anti-greenhouse effect of stratospheric methane.
\item[$\bullet$] H$_2$, through Collision Induced Absorptions, is an extremely powerful greenhouse gas that could potentially warm the surface of TRAPPIST-1h well above the melting point of water \citep{Stevenson:1999,Pierrehumbert:2011,Luger2017}. However, given the small size of the planet, and given the preliminary results of transit spectroscopy with HST \citep{Dewit2016}, the possibility of an H$_2$-rich atmosphere around TRAPPIST-1h seems unlikely.
\end{enumerate}

\section{Conclusions}

In this paper we have used sophisticated numerical models (a N-body code and a Global Climate Model) 
to better constrain the nature of the TRAPPIST-1 planets. The main conclusions of our paper are summarized below:

\subsubsection*{Tidal dynamics constraints}

We showed that, given the low eccentricities derived from our N-body numerical simulations, 
the seven planets of the TRAPPIST-1 system are very likely in synchronous rotation today, 
with one side permanently facing their ultra-cool host star TRAPPIST-1, and one side in the permanent darkness.

Using the same N-body simulations, we also showed that tidal heating is expected to be the dominant process 
of internal heating for the three inner planets of the system (TRAPPIST-1b, c and d). Tidal heating could play a significant role on 
TRAPPIST-1e (given that TRAPPIST-1 seems to be an old system and that radiogenic heating should have decreased), 
but should have a much less pronounced effect on the three outer planets (TRAPPIST-1f, g and h).

\subsubsection*{Climate diversity constraints}
Assuming that the TRAPPIST-1 planets are all in synchronous rotation today, we detail below 
the main conclusions of our paper regarding the possible climates of TRAPPIST-1 planets:


\paragraph{$\bullet$ Airless planets should remain airless.}

\medskip

TRAPPIST-1 planets are exposed to X/UV radiation and stellar wind atmospheric erosion, and could have lost their atmosphere earlier in their history. 
We showed that planets that - at some point - completely lost their atmosphere are more likely to remain airless. 
\begin{enumerate}
    \item Planets that have a low internal heat flux (e.g. TRAPPIST-1e, f, g and h) have to accumulate very large quantities of volatiles on their 
nightside before a runaway greenhouse process re-forms a global atmosphere.
    \item Planets that have a large internal heat flux (e.g. TRAPPIST-1b, c and d) would 
struggle to store and protect volatiles located on their nightside. 
The warmer temperature of the nightside should be responsible for the formation of a residual atmosphere that would be exposed to atmospheric escape.
\end{enumerate}

\medskip

However, both TTV analysis of the planets and the compact, resonant architecture of the system suggest that each of the 
TRAPPIST-1 planets could still be endowed with various volatiles today. Assuming that the four TRAPPIST-1 outer planets (e, f, g and h) 
were able to retain various volatiles in their atmosphere, surface or subsurface, we summarize the last part of our results below: 

\paragraph{$\bullet$ Background atmospheres are stable regarding atmospheric collapse.}

\medskip

TRAPPIST-1 planets are all highly resistant to complete atmospheric collapse of N$_2$ or any other background gas (CO, O$_2$). 
Around 10~millibar of N$_2$ or any other background gas should suffice to avoid surface condensation on the nightside of the TRAPPIST-1 planets. 
This is an essential property, because background gases can prevent the more volatile species (CO$_2$, NH$_3$, etc.) from collapsing on the nightside.

\paragraph{$\bullet$ CO$_2$-dominated atmospheres are sensitive to atmospheric collapse and gravitational burial.}
If TRAPPIST-1e, f, and g outer planets have a CO$_2$-dominated atmosphere, this atmosphere must be very thick.
\begin{enumerate}
    \item Thin CO$_2$ atmospheres would collapse permanently on the nightside of the planets. For example, 
a Mars-like atmosphere would be unstable on TRAPPIST-1e, f, g and h.
    \item Thick (multi-bars) CO$_2$ atmospheres are found stable, thanks to an efficient greenhouse warming and heat redistribution. 
For example, a Venus-like atmosphere would be stable on TRAPPIST-1e, f and g. 
Note however that TRAPPIST-1h is beyond the CO$_2$ condensation limit.
\end{enumerate}
If CO$_2$ somehow starts to condense on the nightside of TRAPPIST-1 outer planets, 
it would form CO$_2$ ice glaciers that would flow toward the substellar region. 
A complete CO$_2$ ice cover is not possible for TRAPPIST-1f and the inner planets 
because they receive an insolation that is greater that the runaway greenhouse threshold for CO$_2$.
A complete CO$_2$ ice cover is found possible on TRAPPIST-1g and h only, 
although the CO$_2$ ice glaciers should be gravitationally unstable and get buried beneath the water ice shell (if present) 
in geologically short timescales. CO$_2$ could be permanently sequestred underneath the water ice cover, in the form of CO$_2$ clathrate hydrates or 
dissolved in a subglacial water ocean.
This makes the presence of surface CO$_2$ ice deposits rather unlikely on water-rich, synchronous planets.

\paragraph{$\bullet$ Sustaining continuously a CH$_4$-rich atmosphere is challenged by photochemical destruction.}
Given TRAPPIST-1 planets large EUV irradiation (at least $\sim$~10$^3~\times$ Titan's flux) 
and the large photodissociation rates that are associated, 
sustaining continuously a CH$_4$-rich (and NH$_3$-rich, by analogy) atmosphere over TRAPPIST-1 lifetime is difficult. 
Calculations of the surface temperatures of the three TRAPPIST-1 outer planets (f, g and h), under a CH$_4$-rich atmosphere, indicate that: 
\begin{enumerate}
    \item their surface (even on the nightside) should be too warm to sustain oceans of methane and/or ethane.
    \item their surface should be to cold to sustain surface liquid water. This is mostly due to the 
anti-greenhouse effect of photochemical hazes and stratospheric methane. The planets could then more likely be covered by water ice.
\end{enumerate}
Photochemical hazes when sedimenting could thus form a surface layer of tholins 
that would progressively thicken - over the age of the TRAPPIST-1 system - above the surface.

\subsubsection*{The habitability of the TRAPPIST-1 system.}
Remarkably, provided a sufficient H$_2$O reservoir is present, TRAPPIST-1e should always sustain surface liquid water, 
at least in the substellar region. This stems from the synchronous rotation coupled to an ideal insolation, and is independent of 
the atmospheric background content (from no atmosphere at all, to a thick atmosphere of hundreds of bars). 
The H$_2$O reservoir should be large enough to avoid trapping on the nightside.

Conversely, TRAPPIST-1f, g and h are unable to sustain surface habitability only with background gases 
(i.e. a rather transparent atmosphere that can ensure the transport of heat and the pressure 
broadening of absorption lines of greenhouse gases).
$\sim$~1~bar of CO$_2$ (respectively $\sim$~5~bars) would be needed to raise the surface temperature above the 
melting point of water on TRAPPIST-1f (resp. g). A thick CH$_4$ atmosphere should be unable to sustain surface habitability on 
TRAPPIST-1f, g and h.

TRAPPIST-1h is unable to sustain surface habitability 
with N$_2$, CO$_2$, CH$_4$, etc. only. This could only be achieved with an unlikely, 
thick H$_2$-dominated atmosphere.

\medskip
\medskip

Future atmospheric exploration of the TRAPPIST-1 system with the James Webb Space Telescope and other forthcoming astronomical 
observatories is extremely promising. TRAPPIST-1 planets are about to become invaluable probes for comparative planetary science outside 
our Solar System and possibly habitability. 
The results of our paper could serve to prepare and then interpret the future observations 
of the TRAPPIST-1 system and analogous. 
The various numerical climate simulations presented in this paper will actually be used 
in follow-up papers to provide the community with synthetic observables (transit spectra, phase curves, and secondary 
eclipses), that should be directly comparable with future JWST observations.
Eventually, we remind the reader that the results of this paper can be applied to any other 
cool Earth-sized planets orbiting in synchronous rotation around any cool to ultra-cool star.

\begin{acknowledgements}
M.T. thanks Tanguy Bertrand, Jan Vatant d'Ollone and S\'ebastien Lebonnois for fruitful 
discussions on Pluto and Titan. M.T. also thanks Kevin Olsen for his editorial feedback. 
M.T. thanks B. Charnay and B. B\'ezard for insightful discussions on the photochemistry of Titan.
On behalf of M. Turbo-King, M.T. thanks warmly his colleagues B.R. Tang (aka Tanguy Bertrand), Z. Habeertable (aka Aymeric Spiga), 
M.C. Chouffe (aka Apurva Oza), B. Exquisit (aka David Dubois) and L. Keg-Beer (aka Laura Kerber), for their limitless inspiration.

By the way, this project has received funding from the European Research Council (ERC) under 
the European Union’s Horizon 2020 research and innovation program (grant agreement No. 679030/WHIPLASH).
\end{acknowledgements}

\bibliographystyle{aa} 
\bibliography{biblio_mt} 

\end{document}